\documentclass[12pt,psf,epsf]{JHEP3}
\usepackage[centertags]{amsmath}
\usepackage{amssymb}
\usepackage{graphicx}
\usepackage{epsfig}

\usepackage{epstopdf}

\newcommand{\mq}{\mathfrak{q}}
\newcommand{\ms}{\mathfrak{s}}

\newcommand{\ma}{\mathfrak{a}}
\newcommand{\mc}{\mathfrak{c}}
\newcommand{\mC}{\mathfrak{C}}
\newcommand{\mA}{\mathfrak{A}}

\newcommand{\mm}{\mathfrak{m}}

\newcommand{\bO}{{\bf\Omega}}

\newcommand{\be}{\begin{equation}}
\newcommand{\ee}{\end{equation}}
\newcommand{\bea}{\begin{eqnarray}}
\newcommand{\eea}{\end{eqnarray}}

\newtheorem{theor}{Theorem}
\newtheorem{lemma}{Lemma}

\newtheorem{Remark}{Remark}

\title{Rolling in the Higgs Model and Elliptic Functions\\
}
\author{ I.Ya. Arefeva, E.V. Piskovskiy, and I.V. Volovich\\  Steklov Mathematical Institute, Gubkin str.~8, 119991,
Moscow,
Russia.}

\abstract {Asymptotic methods in nonlinear dynamics are  used
 to improve  perturbation theory results in the oscillations  regime. However, for some problems of nonlinear dynamics, particularly in
the case of Higgs (Duffing)  equation and the Friedmann cosmological
equations,  not only small oscillations   regime is of interest but
also the regime of rolling (climbing), more precisely the rolling
from a top (climbing to a top). In the Friedman cosmology, where
the slow rolling
 regime is often used,  the  rolling from a top (not necessary slow)
is of interest too.

In the present work a method for approximate solution to the Higgs
equation  in the  rolling   regime is presented. It is shown that in
order to improve perturbation theory in the rolling regime  turns
out to be effective not to use an expansion in trigonometric
functions as it is done in case of small oscillations but use
expansions in hyperbolic functions instead. This  regime
 is investigated using the representation of the
solution  in terms of elliptic functions. An accuracy of the
corresponding approximation is estimated.
}

\keywords{oscillations, rolling  regime, climbing regime, Higgs model}
\begin{document}
\newpage
\section{Introduction}

Known asymptotic methods in nonlinear dynamics such as the
Krylov-Bogoliubov averaging method, the Lyapunov method of slow
varying parameters, the Poincare and the Van der Pol methods, and KAM-theory
are used to improve perturbation theory results when describing
small oscillations, see \cite{Kry-Bog,Bogol-Mitrop,
Arnold-Neist-Kozlov, Koz-3}. Even the term itself - nonlinear
mechanics - is frequently considered as a synonym to nonlinear
oscillations theory, see \cite{Kry-Bog}.

However, in case of the Higgs equation in field theory \cite{VR},
the Friedman equation in cosmology \cite{Linde,Mbook,Rubakov} and some
other problems of nonlinear dynamics not only the regime of small
oscillations is of interest but also the rolling regime (see for
example \cite{AV11,ABG} and references within).

Small perturbations of  periodical motions are usually
considered as perturbations  to the linear equation of the harmonic
oscillator:
\be
  \label{osc1}
  \ddot{q}+\omega^2 q=0,
\ee where  $q=q(t)$ is a  real-valued function of time $t$ and
$\omega > 0$.

The rolling (climbing) regime by definition is a small perturbation of the following linear equation:

\be
  \label{rha}
  \ddot{q}-\mu^2 q=0,
\ee
where $\mu > 0$.

In the present work a method to study  nonlinear  systems in the
rolling  regime is presented. This method can be considered as an
hyperbolic analogue of the averaging method \cite{Bogol-Mitrop}. We consider it
 (as an example) in case of the Higgs equation
 \be
\label{l-EOM-m} \ddot q(t) -\mu^2 q(t) =-\epsilon \,q^3(t),\,\,\,\mu
>0,\,\,\epsilon>0,
\ee
where $\epsilon$ is small coupling constant.
  The method gives
approximate solutions  using expansions of the exact solution in
hyperbolic functions.
The following approximate solution is found in Theorem 1 in Sect. 7.3:
\be \label{M-q2-1}
q_{appr}(t)=A\sinh \biggl(t \mu\biggl(1+\frac38\frac{\epsilon A^2}{\mu^2}\biggr)\biggr)
-\frac1{32}\frac{\epsilon^2 A^3}{\mu^4}\sinh(3t \mu) \,,
\ee
where $A $ is an arbitrary constant. One has the estimation
\be \label{M-q2-1-5}
|q(t)-q_{appr}(t)|\leq C\epsilon^{2-\sigma}\,,\,\,\,0<\sigma < 3/4,
\ee
which is valid for time in the interval
\be \label{M-q2-1-2}
0\leq t\leq c_1\log\frac{c_2}{\epsilon}\,
\ee
that expanding with the vanishing $\epsilon$.

It is well-known that solutions of top
 simple equations of nonlinear dynamics can be written  in terms of elliptic functions. Correspondence
between these exact solutions and  solutions obtained by means of asymptotic
averaging method is
considered in the book  \cite{Bogol-Mitrop}, where an equation for
anharmonic oscillator with real frequency is investigated
\bea
\label{intr-1-m}
  \ddot{q}(t)+\omega^2q(t)=-\epsilon q^3(t),\,\,\omega>0,\,\,\epsilon>0\,.
\eea
Numerical comparison of approximated and
exact solutions  of (\ref{intr-1-m}) is presented in \cite{Bogol-Mitrop}.

In the present work, unlike \cite{Bogol-Mitrop},
equation (\ref{l-EOM-m}) with the imaginary frequency  is considered.
It has also  a representation in terms of
elliptic functions.  The rolling regime for solutions to this equations is considered.
Equations
(\ref{intr-1-m}) and (\ref{l-EOM-m}) are connected
through change
$\omega=i\mu$. However, the averaging method is not applicable to
study  the rolling    regime for solutions to (\ref{l-EOM-m}).  We use
another method that consists in  expansion  of solutions in terms of hyperbolic functions.\footnote {It
would be interesting to consider possible correspondence of the
method of expansion in hyperbolic functions used in the present work
with small $t$ with the first Lyapunov method development
\cite{Koz-3} that is effective with $t\to\infty$.}
 Applications of approximate
solutions in functional mechanics are considered in \cite{IVF,IV10,PV}.

It is well known that by a  shift at a constant
$q(t)=x(t)+\mu/\sqrt{\epsilon},$ the equation (\ref{l-EOM-m}) is converted into the
 equation describing nonlinear oscillations
\be
\label{3-EOM} \ddot x
+2\mu^2 x =-\epsilon \,x^3-3\sqrt{\epsilon}\mu \,x^2.
\ee
However this transformation does not allow to investigate the
rolling mode more effectively then explicit consideration of the initial
Higgs equation.

The paper is organized as follows. In Sect.2 we discuss a
hyperbolic analogue of the averaging method. In Sect.3 solutions to
the Higgs equation in terms of elliptic functions is presented.
Expansions of elliptic functions in
series of hyperbolic functions with
 error estimations for
n-mode approximations are presented in Sect.4. A rearrangement
of the expansion of the elliptic function
in a form  suitable to relate them with the series obtained via
the hyperbolic analogue of the averaging method is presented in Sect.5.
The equivalence
 of these two series is noted  in Sect.5.
Estimations of  precision of hyperbolic
approximations are presented in Sect.6.  In  Appendix
proofs of Lemma 1 and Theorem 2 are presented in details.

The paper is based on \cite{AVP,AV}.

\newpage
\section{Hyperbolic analogue of the averaging method}

The simplest  equation considered in nonlinear dynamics has the following form:

\be \label{EOM} \ddot{q}+\omega^2q=\epsilon f(q,\dot{q}),
\ee where $q=q(t)$ is a real-valued function of time $t$, frequency
$\omega$ is a positive number, $\epsilon$ is a small parameter.

In applications the following equation is also considered
\be
\label{TEOM}\ddot{q}-\mu^2q=\epsilon f(q,\dot{q}), \ee where
$\mu^2>0.$ The change of variable $\omega=i\mu$ transforms (\ref{EOM})
into the latter expression. This equation is called an equation of rolling.
For instance, the Higgs equation \cite{VR} and the  tachyon field in the Friedman
cosmology also have this form \cite{ABG}.

By analogy with the Krylov-Bogoliubov expansion method, used to solve equation
(\ref{EOM})
describing oscillations, we will look for a general solution to equation (\ref{TEOM})
describing rolling in the form of the following expansion
\be
  \label{rollr}q=a \sinh \psi+\epsilon u_1(a,\psi) +\epsilon^2
  u_2(a,\psi)+...\; .
\ee
Here $a,\psi$ are functions of time that are determined by the differential equations
\bea
  \label{rollr2}\dot{a}=\epsilon A(a,\epsilon)=\epsilon
  A_1(a)+\epsilon^2 A_2(a)+...\,,\\
  \dot{\psi}=\mu+\epsilon B(a,\epsilon)=\mu+\epsilon
  B_1(a)+\epsilon^2B_2(a)+...\,.
\eea
 Let us consider {\it the following } equation
\be
\label{l-EOM}
  \ddot q(t) -\mu^2 q(t) =-\epsilon \,q^3(t),\,\,\,\mu >0,\,\,\epsilon>0.
\ee
We look for  a solution of the form
\be
  \label{M-q1} q(t)=\mA\sinh(t \mu\mm) +  ...
\ee
and define parameters $\mA, \mm$ in a way to keep the functional form of the solution after substitution in the equation  (\ref{l-EOM}).
We get on the right-hand side of (\ref{l-EOM}):
\bea
  \label{eq4}
  -\epsilon(\mA\sinh(t \mu\mm) +....)^3 =-\epsilon\left(\frac14\mA^3\sinh(3t \mu\mm) +3\sinh(t\mu\mm)) + ....\right).
\eea
On the left-hand side of (\ref{l-EOM}) we obtain:
\bea
(\partial ^2-\mu^2)(\mA\sinh(t \mu\mm) + ....) = \mA \mu (\mm
^2-1)\sinh(t \mu\mm) +....\;. \eea
We see that in order to get an equality we have to equate
\be \label{shift-omega} \mA
  \mu ^2(1-\mm^2)=-\epsilon\frac34\mA ^3.
\ee
Therefore it follows that
\be
\mm^{2} =1+\frac34\frac{\epsilon\mA^2}{\mu^2}.\ee

In order to keep  the form of the approximation in the right-hand side and the left-hand sides of (\ref{l-EOM})  one has to take the next term of the approximation  in the form
\be \label{M-q2}
q(t)=\mA\sinh\biggl(t \mu\biggl(1+\frac38\frac{\epsilon\mA^2}{\mu^2}\biggr)\biggr)
+\mA_1\sinh(3t \mu)  +...\,.
\ee Taking into account (\ref{eq4}) we get
\be
\label{M-q2-3} \mA_1=-\frac1{32}\frac{\epsilon^2\mA^3}{\mu^4}.
\ee
Approximation (\ref{M-q2}), (\ref{M-q2-3}) is considered in \cite{AV11}.

The described procedure can be carried on and it turns out that
for equation (\ref{l-EOM}) with initial conditions $\dot q(0)=0$
 the procedure leads to the representation of the solution  in the form of an infinite sum of $n$-mode terms
\be \label{as-t} q(t)=\sum^\infty_{n=0}\ms_n
(t),
\ee
where
\bea \label{as-t0} \ms _0(t)&=&\mA\sinh\left(\mu\mm\,
t\right),\,\,\,\,\,\,\,\,\,\,\,\,\,\,\,
\,\,\,\,\,\,\,\,\,\,\,\,\,\,\,\,\,\,\mm=\mm(\lambda)\,,\\
\label{as-tn}\ms _n(t)&=&\mA \,\ma_n(\lambda)\sinh\left((2n+1)
\mu\mm\, t\right),\,\,\,\,\,\,\,\,\,\,\,\,\,\,n=1,...\; .
\eea
Here $\mm \left(\lambda\right)$ and $\ma_n\left(\lambda\right)$ are power series of dimensionless parameter
\be\label{lambda}
\lambda=\frac{\mA^2\epsilon}{\mu^2},
\ee
i.e.
\bea \mm\,
(\lambda)&=&1+\mm\,_1\lambda
+...+\mm\,_n\,\lambda^n+...,\\
\ma_n\left(\lambda \right)&=&\ma_{n0}\,\lambda^n
+\ma_{n\,1}\,\lambda^{n+1}+... +\ma_{n\,n}\,\lambda ^{2n}+...\;.
\eea
The solution depends on parameters $\epsilon, \mu$, and it is specified by the only parameter $\mA$.
Within this notation the first terms of (\ref{as-t}) are written in the form
\bea
\label{as-t-hyp} q(t)&=&\mA\sinh\left(\mu\biggl(1+\frac
{3\mA^2}{8\mu^2}\epsilon - \dfrac{15 \mA^4}{2^8
\mu^4}\epsilon^2\biggr)t\right)\\ \nonumber &-& \mA\left(\frac{\epsilon
\mA^2}{32\mu^2}-\frac{21\epsilon^2
\mA^4}{1024\mu^4}\right)\sinh\left(3\mu\biggl(1+\frac
{3\mA^2}{8\mu^2}\epsilon \biggr)t\right)+\frac{21\epsilon^2
\mA^5}{1024\mu^4}\sinh(5\mu t) +\,...
\eea
This series can be  formally obtained from
 the result of the averaging method (see (\ref{as-1}) below) by the change $\omega \rightarrow i \mu, \,A \rightarrow
-i\mA. $

For the Higgs equation with friction:
\be \label{friction} (\partial
^2+2\epsilon h\partial -\omega^2)q+\epsilon q^3=0,\,\,\,\,\,q(0)=0
\ee
an approximate solution has the form
\bea q(t)&=&\mA\,e^{-\epsilon h
t}\sinh\left(\mu\biggl(t-\frac{3\mA^2}{16\mu^2 h}(1-e^{-2\epsilon h
t})\biggr)\right)\nonumber
\\
&-&\epsilon \frac{\mA^3}{32\mu ^2} e^{-3\epsilon h t}\sinh\left(3\mu
\biggl(t-\frac{3\mA^2(1-e^{-2\epsilon h t})} {16\mu ^2 h}\biggr)\right). \eea
The method developed in this work can also be used to investigate a rolling mode (not necessarily slow one) for the Friedman equation either in local or in nonlocal cosmology, see \cite{Rubakov,AV11}:
\be
\label{repr21}
  \ddot{q}+\epsilon_1\sqrt{\frac{\dot{q}^2}{2}\pm\omega^2\frac{q^2}{2}
+\epsilon_2 V(q)}\,\pm\omega^2 q+\epsilon_2 V^{\prime}(q)=0.
\ee

\section{Higgs equation and elliptic functions}
\subsection{Solution to the Duffing equation}
In this section well-known solutions to some nonlinear equations  are presented.

As it is known the equation of the form
$$
\ddot{q}+V^{\prime}(q)=0
$$
has the first integral:
$$
E=\frac12\dot{q}^2+V(q)
$$
and solution to the equation can be presented in the form of quadrature
$$
t=\int\frac{d q}{\sqrt{2(E-V(q))}}.
$$

If $V(q)$ is a quartic polynomial, then the latter integral can be represented in terms of elliptic functions \cite{Zur}, \cite{Red}, \cite{Akh}.

It follows that the solution to the equation for the  anharmonic oscillator (the Duffing equation)
\bea
\label{intr-1}
  \ddot{q}(t)+\omega^2q(t)&=&-\epsilon q^3(t),\,\,\omega>0,\,\,\epsilon>0
\eea has the representation \bea
 \label{sol-1}
 q(t)=a \,{\rm {cn}}(\Omega t+b,k),
 \eea
where $\,{\rm {cn}}(u,k)$ is the elliptic cosine function of argument $u$ and
modulus  $k$, $a$ is the amplitude and $b$ is the phase.  The frequency $\Omega$ and
the modulus of  the elliptic function $k$ are obtained from parameters
$\omega, \epsilon$ of the equation  and depend on the  amplitude $a$:
\begin{align}
  \Omega = \sqrt{\omega^2 + \epsilon a^2},\qquad k = \sqrt{\dfrac{\epsilon a^2}
  {2(\omega^2 + \epsilon ^2)}}.
\end{align}

The function ${\rm {cn}}(\Omega t + b,k)$ can be expanded in trigonometric series
\begin{equation}
  {\rm {cn}}(\Omega t + b,k) = \dfrac{2 \pi}{k {\rm {\bf K}}}
  \sum\limits^{\infty}_{n=1} \dfrac{\mq^{n-1/2}}{1+\mq^{2n-1}}
  \cos\biggl((2n - 1)\dfrac{\pi (\Omega t + b)}{2 {\rm {\bf K}}}\biggr),
\end{equation}
where
\begin{equation}
\mq = e^{-\pi\frac{{\rm {\bf K'}}}{{\rm {\bf K}}}}, \qquad \mq' =
e^{-\pi\frac{{\rm {\bf K}}}{{\rm {\bf K'}}}}.
\end{equation}
Here ${\rm {\bf K}}={\rm {\bf K}}(k)$ is complete
elliptic integral of the first kind and ${\rm {\bf K'}}={\rm {\bf K}}(k')$, where $k^2 + k'^2 = 1$ \cite{Zur}, \cite{Red}, \cite{Akh}. In particular, with $b=-{\rm {\bf K}}$
one obtains:
\begin{equation}
  {\rm {cn}}(\Omega t - {\rm {\bf K}},k) = \dfrac{2 \pi}{k {\rm {\bf K}}}\sum\limits^{\infty}_{n=1}\dfrac{\mq^{n-1/2}}{1+\mq^{2n-1}}\sin\biggl((2n-1)\dfrac{\pi \Omega t}{2 {\rm {\bf K}}}\biggr).
\end{equation}
In case of small $\epsilon$ one can obtain
an asymptotic expansion from this representation

\be
\label{as-1}
  q(t)=A\sin\left(\omega \biggl(1+\epsilon \frac {3A^2}{
8\omega ^2}\biggr)t\right) -\epsilon
\frac{A^3}{32\omega^2}\sin\left(3\omega \biggl(1+\epsilon \frac {3A^2}{
8\omega ^2}\biggr)t\right)+...,
\ee
here $A$ is an arbitrary real parameter.
One can also obtain this expression
 by means of the Krylov-Bogoliubov averaging method \cite{Bogol-Mitrop}.

\subsection{Solution to the Higgs equation}
Solution to the equation
\be
\label{l-EOMm}
  \ddot q -\mu^2 q =-\epsilon \,q^3,\,\,\,\,\,\epsilon>0
\ee
can belong to one of the three types depending on initial conditions
\be
q (0)=q_0,\,\,\,\,\,\dot q(0)=v_0.\label{ID}
\ee
A type of the solution is determined by the energy
\be \label{Energy}
E=\frac12 v_0^2-\frac12\mu^2q_0^2+\frac14 \epsilon  q_0^4.
\ee
Solutions have one of the following forms
\begin{figure}[h!]
\centering
\includegraphics[width=35mm]{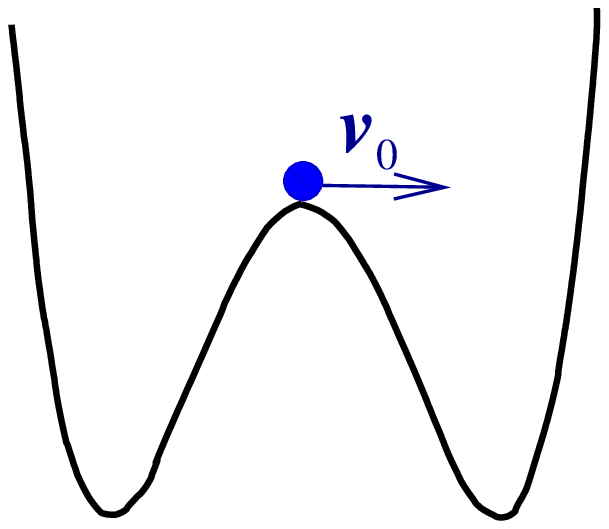}A.$\,\,\,\,\,\,\,\,\,\,\,$
\includegraphics[width=35mm]{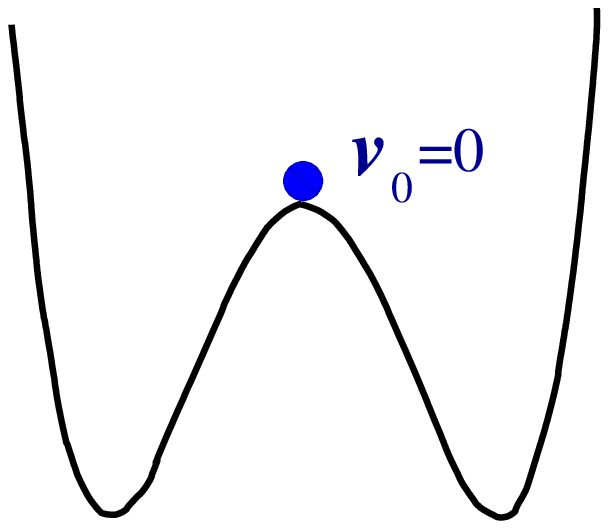}B.$\,\,\,\,\,\,\,\,\,\,\,$
\includegraphics[width=35mm]{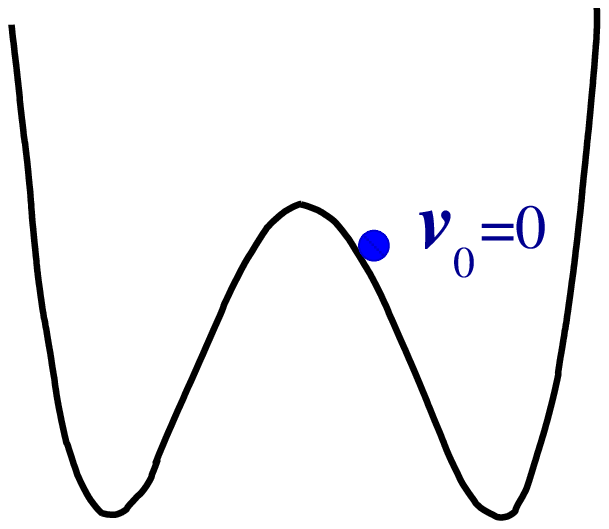}C.
\caption{Potential energy and initial conditions. A. The initial condition corresponding to a motion in two holes, $E>0$; 
B. The initial condition, corresponding to a motion in two holes with infinite period, $E=0$; 
C. The initial condition corresponding to a motion in one hole, $E<0$.
} \label{pot}
\end{figure}

\begin{itemize}
\item $E>0$; the two-holes solutions (Fig 1A).
\bea \label{2holes-g}
q(t)&=&a\,{\rm {cn}}(\Omega t+b,k)\\
a^2&=&\frac{\mu^2}{\epsilon}\left(1+\sqrt{1+\frac{4\epsilon E}{\mu^4}}\right)\label{A-E-tm}\\
\Omega^2&=&\mu^2\sqrt{1+\frac{4\epsilon E}{\mu^4}}\label{O-E-tm}\\
k^2&=&\frac12+\frac12\,\frac{1}{\sqrt{1+\frac{4\epsilon E}{\mu^4}}}\label{k_E_pos}\\
k^{\prime 2}&=& \frac12-\frac12\,\frac{1}{\sqrt{1+\frac{4\epsilon
E}{\mu^4}}}\label{kp-E-tm}\eea

The parameter $b$ is determined  from the condition
\bea q(0) &=& a\,{\rm
{cn}}(b,k).\,\,\,\,\,\,\,\,\,\,\,\,\,\,\,\,\,\,\,\,\,\,\,\,\,\,\,\,\,
\eea
We will discuss the case with  the initial condition $q(0)=0$. In this case
\be b=-{\rm {\bf K}}
\ee
and
\bea
\label{2holes-pos}
q(t)&=&a\,{\rm {cn}}\,(\Omega t -{\rm {\bf K}},k).
\nonumber \eea
For this solution
\bea
\label{2holes-pos}
v\,=\, \dot q(0)&=&
\frac{a\mu  k^{\prime }
}{\sqrt{1-2k^{\prime 2}}}.
\nonumber \eea
The energy $E$ is expressed in terms of parameter $k'$,

\begin{equation}
 \frac{ E \epsilon}{\mu ^4} = \dfrac {\mu^4 k'^2
 \left(1 - k'^2 \right) }{\, \left(2\,k'^2-1 \right)^2 }.
\end{equation}
We also note the dependency between $a$, $\epsilon$ and $k^\prime$
\be \label{ak1}
a^2=\frac{\mu^2}{\epsilon}\left(1+\frac{1}{1-2k^{\prime
2}}\right)\ee
and  $ \Omega$ with $\mu$ and $k'$
\be
\label{Omega-m} \Omega^2 =\frac{\mu^2
}{1-2k^{\prime 2}}.
\ee

\item $E=0$; the two-holes solution
with period $T=\infty$

\be \label{ezero}
q(t)=\mu\,\sqrt{\frac{2}{\epsilon}}\,\,\frac{1}{\cosh(t\mu+b)}\ee
Parameter $b$ is determined with the initial condition
\be
q(0)=\mu\,\sqrt{\frac{2}{\epsilon}}\,\,\frac{1}{\cosh(b)}
\ee

\item $E<0$; the one-hole solution, for instance, the solution in the right one
\bea \label{1holes-g}
q(t)&=&a\,{\rm {dn}}\,(\Omega t +b,k)\\
a^2&=&\frac{\mu^2}{\epsilon}\left(1+\sqrt{1+\frac{4\epsilon E}{\mu^4}}\right)\label{a_E_neg}\\
\Omega^2&=& a^2\frac{\epsilon}{2}=\frac{\mu^2}{2}\left(1+\sqrt{1+\frac{4\epsilon E}{\mu^4}}\right)\\
k^2&=&2-\frac{2\mu^2}{\epsilon
a^2}=2-\frac{2}{\left(1+\sqrt{1+\frac{4\epsilon
E}{\mu^4}}\right)}\label{k_E_neg} \eea
Parameter $b$ is determined  with the relation
\bea q(0) &=& a\,{\rm
{dn}}(b,k)\,\,\,\,\,\,\,\,\,\,\,\,\,\,\,\,\,\,\,
\,\,\,\,\,\,\,\,\,\,\,\,\,\,\,\,\,\,\,\,\,\,\,\,\,\,\,\,\,\,\,\,\,\,\,\,\,\eea

In the case with the initial data of the special form $v_0=0$ we have
\bea \label{2holes-neg} q(t)&=&a\,{\rm {dn}}\,(\Omega t -{\rm {\bf
K}},k). \eea
The rolling mode corresponds to motion in the vicinity of the turning point, that is closest to the top of the potential (see Fig.\ref{pot} C).
\end{itemize}

\section{Elliptic functions expansion}

We consider solution  (\ref{2holes-g}).
This solution is periodic with period $T = 4{\rm {\bf K}}/\Omega$,
here ${\rm {\bf K}}={\rm {\bf K}}(k)$. We show that for
$t$, such that $\lvert t \rvert < T / 2$, the solution can be expanded in hyperbolic series (\ref{as-t-hyp}).
We also show that the first terms of the expansion
(\ref{as-t}) do represent an approximation to the exact solution
(\ref{2holes-g}) with $b=-{\rm {\bf K}}$.

This remark is of importance because in case of more complicated potentials
 we can explicitly apply the hyperbolic analogue of the Krylov-Bogoliubov method,
 while the exact solutions remain unknown (see, for instance, the problem with
 friction, (\ref{friction})).

We need an expansion of elliptic functions about the point that is equal to minus half-period, i.e. ${\rm
{cn}}(u-{\rm {\bf {K}}},k) $, for small $k^\prime$. In case of small $k'$
 ${\rm {\bf {K}}}\to \infty$ as
\be
 {\rm {\bf K}}\approx\ln \frac 4{k'} +\frac{k^{\prime 2}}{4} \ln\frac 4{k^{\prime 2}} -\frac{k^{\prime 2}}{4} +...\;.
\ee
According to (\ref{kp-E-tm}) if $\epsilon \to 0$, then $k' \to 0$ and vice versa
\be k^{\prime }={\frac
{\sqrt{E\epsilon}}{{\mu}^{2}}}\, \left(1-\frac32\,{\frac
{E\,\epsilon}{{\mu}^{4}}}+ {\cal O}(\epsilon^2)\right).
\ee
Elliptic functions summation formula together with
\be
  {\rm{sn}}({\rm {\bf {K}}})=1, \,\,\,\,\,\,\,{\rm {cn}}({\rm {\bf
{K}}})=0, \,\,\,\,\,\,\,{\rm {dn}}({\rm {\bf {K}}})=k^{\prime }
\ee
gives
\bea
  \nonumber{\rm {cn}}(u-{\rm {\bf {K}}}) &=&\frac{{\rm
{cn}}(u){\rm {cn}}( {\rm {\bf {K}}})+
{\rm {sn}}(u){\rm {sn}}( {\rm {\bf {K}}}){\rm {dn}}(u){\rm {dn}}( {\rm {\bf {K}}})}{1-k^2 {\rm {sn}}^2(u){\rm {sn}}^2( {\rm {\bf {K}}})}\\
&=& \frac{ {\rm {sn}}(u){\rm {dn}}(u)k^\prime }{1-k^2 {\rm
{sn}}^2(u)} = \label{cn-K} \frac{ {\rm {sn}}(u){\rm {dn}}(u)k^\prime
}{{\rm {dn}}^2(u)}=k^\prime\frac{ {\rm {sn}}(u) }{{\rm {dn}}(u)}.
\eea

\subsection{Perturbation theory}
In this section we investigate an expansion of the solution to (\ref{l-EOM}) in
terms of the Jacobi elliptic functions in the case of small parameter $k^{\prime}$ (i.e. $\epsilon$) and compare the first terms of the series with the exact solution.

For small $k^\prime$ the following expansions take place \cite{Red}, p.386, equations  16.15.2-16.15-3,
\bea
{\rm {sn}}(u,k) &=& \tanh u +\frac14
k^{\prime 2}\left(\sinh u\cosh u -u\right)\frac{1}{\cosh^2
u}+... \;,\\
{\rm {dn}}(u,k) &=& \frac1{\cosh u} +\frac14 k^{\prime
2}\left(\sinh u\cosh u +u\right)\frac{\sinh u}{\cosh^2 u}+...\;.
\eea
Consequently, for small $k^\prime$ we have:

\bea \frac{{\rm {sn}}(u,k)}{{\rm {dn}}(u,k)}&=& \sinh u  +\frac14
k^{\prime 2}\left(\sinh u -\sinh^2 u\right)\nonumber\\&-&\frac14 k^{\prime
2}u\left(\frac{1}{\cosh u}-\tanh u \right)+...\eea
and we get the following expansion
\bea
\label{as-tach-cn-p} {\rm {cn}}(u-{\rm {\bf {K}}},k)&=&k' \mc_0(u)+k^{\prime 3}\mc_1(u)+...\;,
\eea
where
\bea
  \mc_0(u)&=&\sinh u\label{pert1},\\
  \mc_1(u)&=&\frac14
           \left(\sinh u -\sinh^2 u\right)- \frac14 u\left(\frac{1}{\cosh u}-\tanh u \right).\label{pert2}
\eea

\begin{figure}[h!]
\centering
\includegraphics[width=35mm]{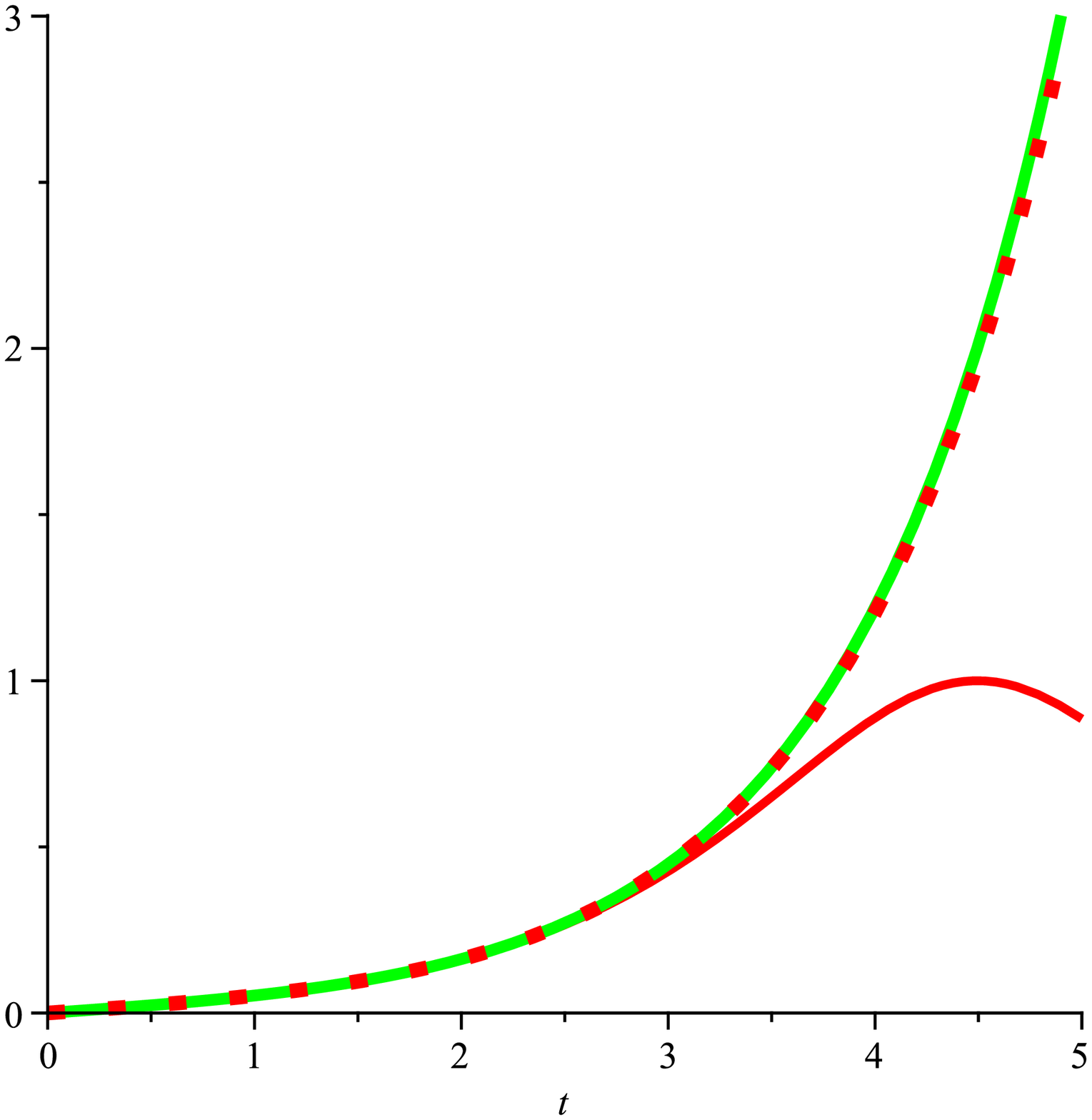} $ A.\,\,\,\,\,\,\,\,\,\,\,\,$
\includegraphics[width=35mm]{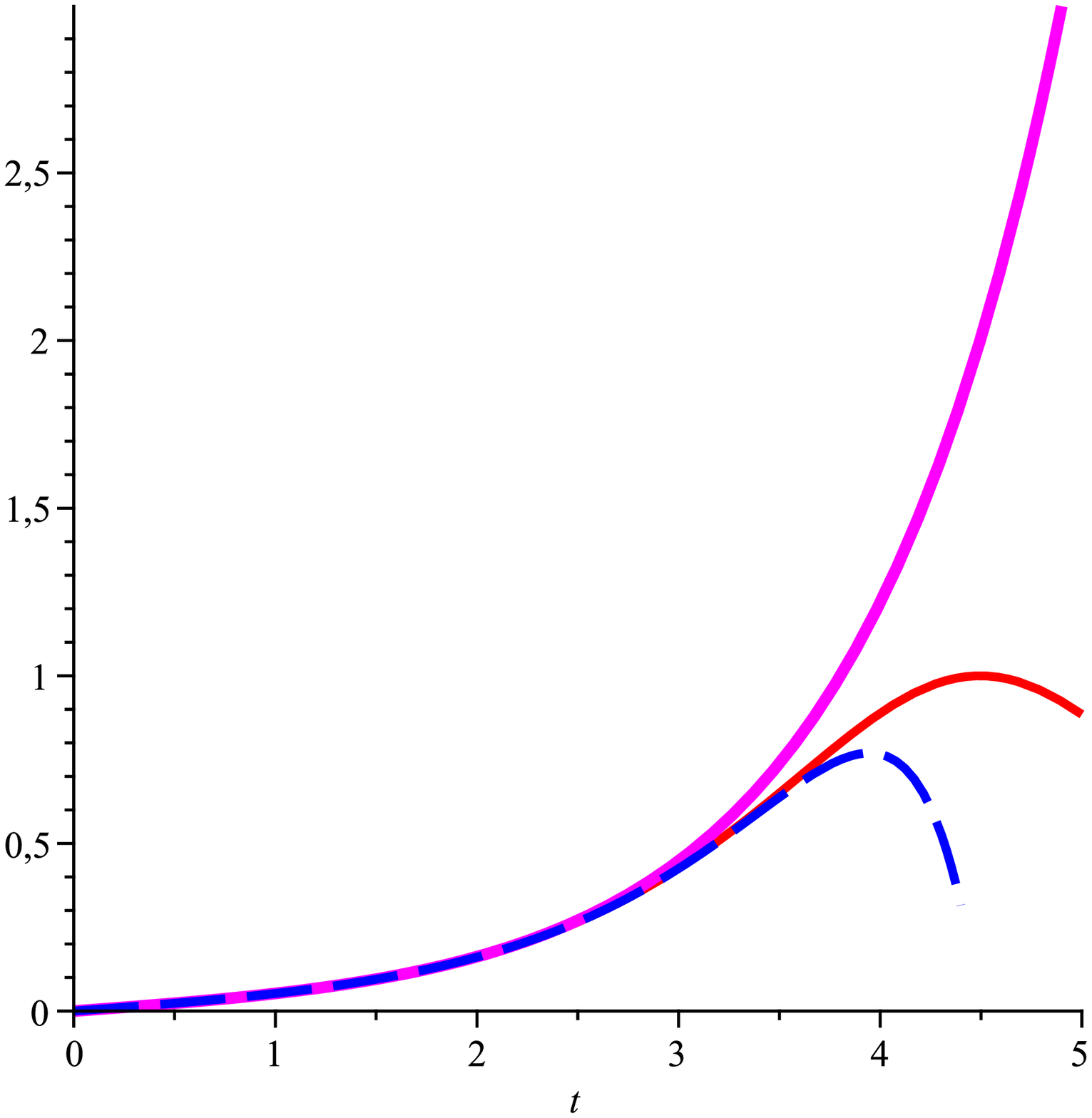}$ B.$\\
\includegraphics[width=35mm]{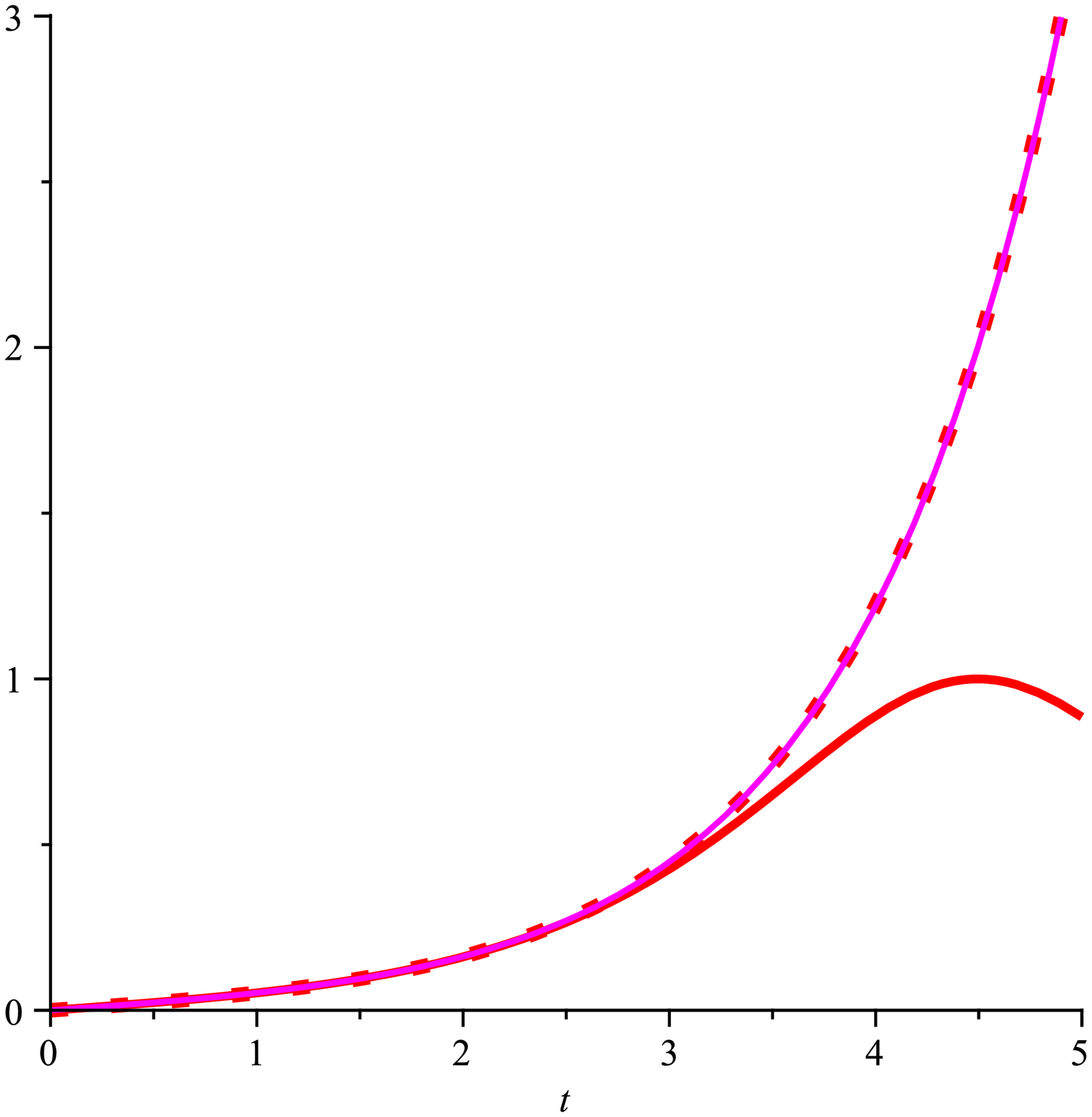} $C.\,\,\,\,\,\,\,\,\,
\,\,\,\,\,$
\includegraphics[width=40mm]{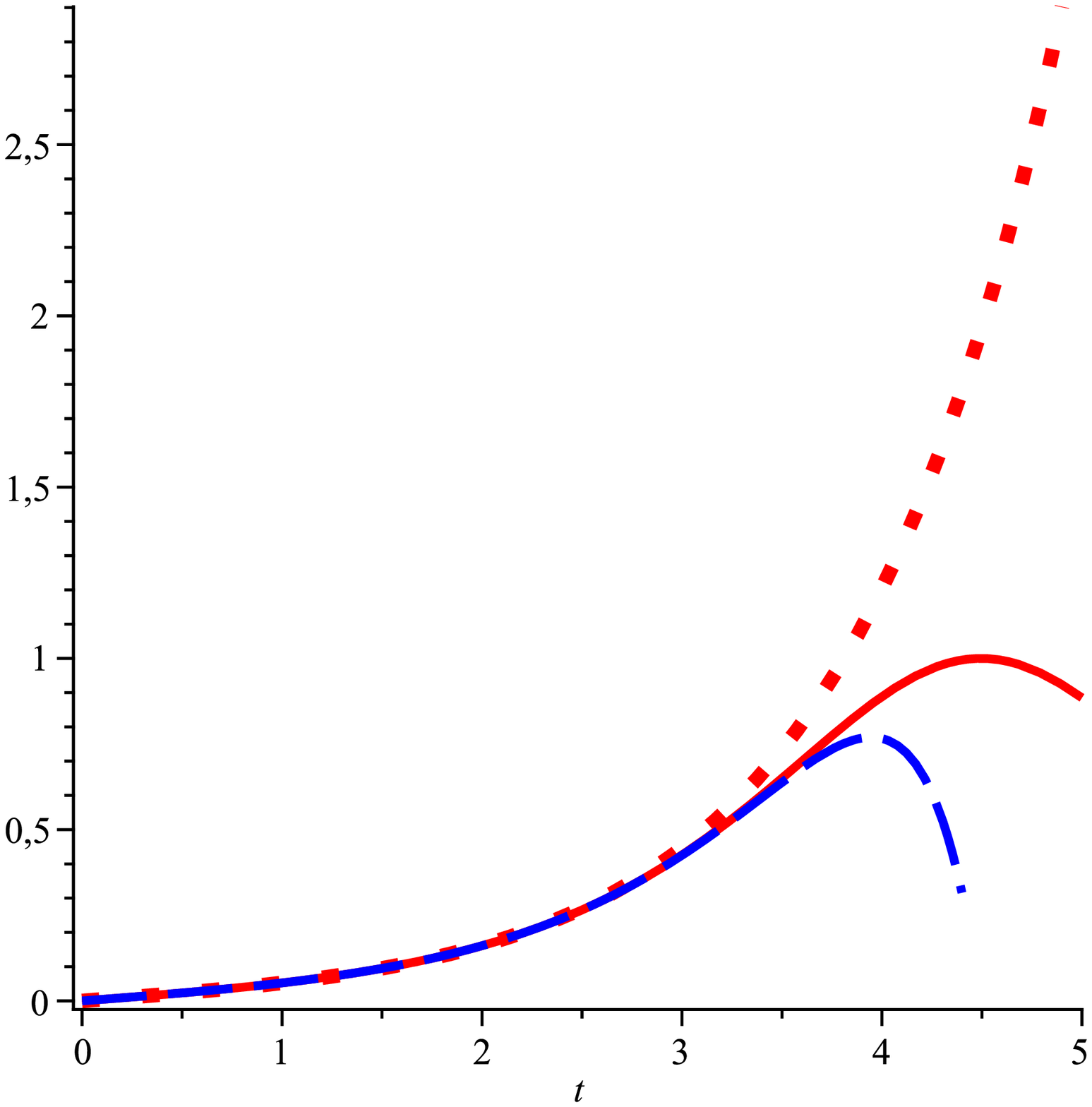} $D.$
\caption{A. A comparison between the first and the second approximations, calculated according to
conventional perturbation theory,
(4.9) 
and
(4.10), 
B. with the first two approximations of the modified perturbation
theory,
C. (4.15) 
and
(4.16), 
D. and with the exact
expression $cn(u-{\rm\bf{K}},k)$, $k=0.999$.}
\label{cn-per}
\end{figure}

Expansion (\ref{pert2}) contains a secular term $k^{\prime 3}u$, however,
unlike trigonometric case, it is dominated by the
term $\frac{1}{4}k^{\prime 3}\sinh^{2} u$ with large $u$.
First two terms of the expansion (\ref{pert2}) are presented in Fig.\ref{cn-per} A.
Modified perturbation theory will be developed using an expansion in hyperbolic
functions in the next subsection. This modified theory gives a better
approximation to the exact solution for large $u$.
It is interesting to mention, that, as it will be
shown in the next subsection in the modified
perturbation theory the term $-\frac{1}{16} k^{\prime 3}\sinh 3u$
contributes greately. Its contribution increases faster
with increase of $u$, than the term from perturbation
theory $\frac{1}{4}k^{\prime 3}\sinh^2 u$.

In the Fig. \ref{cn-per} the comparison of the first two terms
of the perturbation
theory with the exact solution is presented.
The first approximation is the upper solid line, the
second one is plotted by the dotted line. The lower thick line
shows the exact solution. It can be seen that the first and the
second approximations of the conventional perturbation theory almost
coincide with each other. The second approximation calculated according
to the conventional perturbation theory does not in fact improve the first one.

\subsection{Modified perturbation theory}

We will show that the function ${\rm {cn}}(u-{\rm {\bf {K}}},k) $
with $|u|<{\rm {\bf {K}}}$ and small $k'$ has the following representation
\be
\label{as-tach-cn} {\rm {cn}}(u-{\rm {\bf {K}}},k)= \sum _0^\infty
(-1)^n{\cal A}_{n} \sinh ( (2n+1)u'),
\ee
where \bea \rho '&=&\frac{\pi
{\rm {\bf {K}}}}{2{\rm {\bf {K}}}'},\,\,\,\,u'=\frac{\pi u}{2{\rm
{\bf {K}}}'} \eea
\bea
{\cal A}_{n}&\equiv& \frac{\pi}{k {\rm {\bf {K}}}'}\frac{1}{\cosh (2n+1)\rho ')}
 \eea
and the series (\ref{as-tach-cn}) converges with
$|\mathfrak{R}u|<{\rm{\bf K}}$ and $-1<k<1$.

In order to prove (\ref{as-tach-cn}) we use (\ref{cn-K}). Note that for $ \frac{ {\rm
{sn}}(u) }{{\rm {dn}}(u)}$ the following representation holds
(see equation (133), p. 85, \cite{Zur}):

\bea\label{cn-dk} \frac{ {\rm {sn}}(u) }{{\rm
{dn}}(u)}&=&\frac{\pi}{k k^\prime{\rm {\bf {K}}}'}
\left\{\frac{\sinh u'}{\cosh \rho '}-\frac{\sinh 3u'}{\cosh 3\rho
'}+ \,\frac{\sinh 5\rho '}{\cosh 5u'}+...\right\} \eea
multiplication of (\ref{cn-dk}) by $k^\prime$ gives (\ref{as-tach-cn}).

We will discuss approximations
 \bea
\label{mod-pert1}
\mC^{(0)}(u)&=&{\cal A}\sinh u',\\
\label{mod-pert2}\mC^{(2)}(u)&=& {\cal
A}\sinh u'-{\cal A}_1\sinh 3u',
\\
\label{mod-pertn}\mC^{(n)}(u)&=& {\cal A}\sinh u'-{\cal A}_1\sinh
3u'+...+(-1)^n{\cal A}_{n}\sinh (2n+1)u'.
\eea
Within our notation ${\cal A}\equiv {\cal A}_0$. We note, that $u'$ and ${\cal A}_{2n+1}, n=0,1,2, $ have the following expansions:
\bea \label{u}
  u' &=&\frac{\pi }{2{\rm {\bf {K}}}'}u\approx(1-\frac{1}{4} k^{\prime 2}+...\,)\,u, \\
  \label{calA}{\cal A}&=& k^\prime+\frac {7}{16}{k^\prime}^{3}+{\frac
  {79}{256}}{k^\prime}^{5 }+O \left( {k^\prime}^{7} \right),\\
  \label{calA1}{\cal A}_1&=&-\biggl({\frac {1}{16}}{k^\prime}^{3}+{\frac
  {1}{16}}{k^\prime}^{5}+
  O \left({k^\prime}^{7} \right) \biggr),\\
  \label{calA2} {\cal A}_3&=&\frac{1}{4096}k^{\prime 7}+{\cal O}
  \left({k^\prime}^{9} \right).
\eea

However, in our notations we do not expand ${\cal A}_i$ and $u'$
 in power series of $k^{\prime 2}$ in formulae (\ref{u}) - (\ref{calA2}).

\begin{figure}[h!]
\centering
\includegraphics[width=100mm]{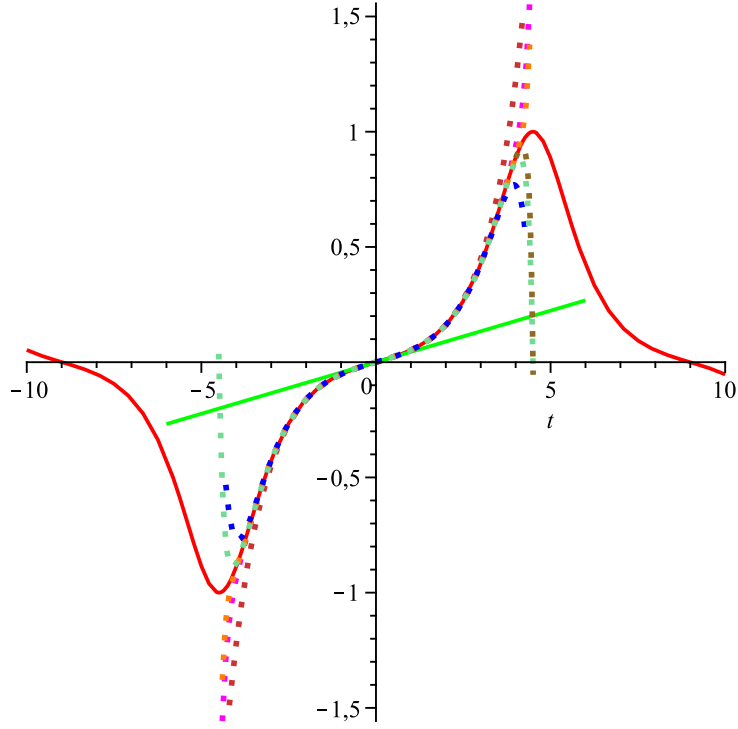} $ A.\,\,\,\,\,\,\,\,\,$
\caption{ (color online) The function of $t$ $cn(t-{\rm\bf{K}},k), k = 0.999$
is shown with the red line. The green line depicts a
linear approximation. The orange, purple and coral dashed lines
depict the first, the third and the fifth approximations calculated
by formula
(4.11) 
respectively dotted lines of cyan, aquamarine and
ochre color depict approximations of the second, the fourth and the
sixth orders calculated by (4.11). 
} \label{apr-cn-m} \end{figure}

Approximations that are truncated series of expansion
(\ref{as-tach-cn}), namely  approximations (\ref{mod-pert1}),
(\ref{mod-pert2}) are shown in Fig.\ref{cn-per}. B. A comparison of
these approximations with ones obtained within the conventional
perturbation theory are presented in Fig.\ref{cn-per}.C and
\ref{cn-per}.D. In Fig.\ref{cn-per}.B the first two approximations
calculated according to the modified perturbation theory are shown.
It can be seen that the second approximation is an improvement of
the first one. A comparison of the first approximations calculated
according to conventional and modified perturbation theories are
presented in Fig.\ref{cn-per}.C. These approximations almost
coincide with each other. As it can be seen in
Fig.\ref{cn-per}.D, the second approximation
calculated according to the conventional perturbation theory and the
second approximation calculated according to the modified
perturbation theory differ significantly.
The second approximation calculated according to
the modified perturbation theory is closer to the exact solution in
comparison with the second approximation obtained according to the
conventional perturbation theory. Next terms of the expansion are
shown in Fig.\ref{apr-cn-m}.

$\,$

\subsection{Bounds for elliptic cosine expansion}
Let us prove the following lemma.
\begin{lemma}\label{L1}
 Let $\mu>0$ and $0<c<1 / 2$, then the following inequality holds:
    \begin{equation}
    \label{estim-m}
        \left|{\rm {cn}}(\Omega t-{\rm {\bf {K}}},k) -
        \mC^{(n-1)}(\Omega t)
 \right| \leq \dfrac{4}{k }(k')^{(2n+1)(1-c)},\,\,n=1,2,...
    \end{equation}
    for all
 \be
  \label{t-rest} 0\leq t\leq\frac{c}{\mu\sqrt 2 }\ln \frac{1}{k'}, \,\,\,\,
    0\leq k'\leq 1/2,
\ee
 here $\mC^{(n)}(u)$ stays for the approximation defined by (\ref{mod-pertn}).
\end{lemma}
\begin{Remark}  Estimation (\ref{estim-m}) is not trivial in comparison to the proposition that the series  (\ref{as-tach-cn}) converges with $|\mathfrak{R}u|<{\rm{\bf
K}}$  and $-1<k<1$. The reason for that is that the estimate (\ref{estim-m}) according to (\ref{t-rest}) allows to consider $t\to \infty$ with $k'\to 0$.
\end{Remark}
{\it Proof} is based on the estimation of the residual term
\bea r _n\equiv\dfrac{\pi}{k {\rm {\bf
{K}}}'} \left\{-\dfrac{\sinh ((2n+1)\rho' u'')}{\cosh ( (2n+1)\rho
')}+ \,\dfrac{\sinh ( (2n+3)\rho' u'')}{\cosh ( (2n+3)\rho
')}+...\right\},  u'' = \dfrac{u}{{\rm {\bf
{K}}}}\eea
by the geometric sequence sum
\be \label{main-dm}d_n\equiv\dfrac{\pi}{k {\rm
  {\bf {K}}}'} \dfrac{e^{-(2n+1)\rho'(1 - u'')}} {1 - e^{-2\rho'(1 -u'')}},
\ee
that is
\be
 |r_n| \leq d_n.\label{main_estmnm} \ee
Taking into account
\be
e^{-2\rho'(1 -
u'')}=\exp\left(-2\dfrac{\pi {\rm {\bf {K}}}}{2{\rm
{\bf {K'}}}}\right)\exp\left(\dfrac{2\pi\Omega t}{2{\rm {\bf
{K'}}}}\right)\ee
the estimation (\ref{main_estmnm}) holds if
\bea
  \label{est-1}
  \exp\left(-\dfrac{\pi {\rm {\bf {K}}}}{2{\rm {\bf{K'}}}}\right)\exp\left(\dfrac{\pi\Omega t}{2{\rm {\bf
  {K'}}}}\right)<1.
\eea
If this condition holds then
\bea
  \label{est-dn}
  d_n\leq \dfrac{4}{k} \biggl(k'\biggr)^{2n+1}e^{(2n+1)\Omega t}<\dfrac{4}{k}(k')^{(2n+1)(1-c)}\to 0\\
   \,\,\,\,for \,\,\,\,k'\to0\,\,\,and\,\,\,\,c<1.
\eea
The condition (\ref{est-1}) is satisfied if there is a restriction on $t$, such that
\be \label{cond-m}
     \biggl(k'\biggr)^2\exp\left(2\Omega t\right) < \frac12 .
\ee
Taking into account the latter limitation and the property of the elliptic functions formulated below the condition (\ref{est-1}) we get
\bea\exp\left(-\dfrac{\pi {\rm {\bf {K}}}}{2{\rm {\bf
{K'}}}}\right)\exp\left(\dfrac{\pi\Omega t}{2{\rm {\bf
{K'}}}}\right)<k'e^{\Omega t} . \eea
The detailed proof can be found in Appendix A.

\section{$n$-mode approximation}
In the present section approximations of exact solutions are presented and
accuracy of these approximations is estimated for small coupling
constant $\epsilon$.

Taking into account the series (\ref{as-tach-cn}) let us consider an approximation of the exact solution (\ref{2holes-pos}).
We get
\be
\label{as-tach-cn-m}
q(t) =a\,{\rm
{cn}}(\Omega t -{\rm {\bf {K}}},k)= \sum_{n=0}^{\infty}s_n(t), \ee
where the following notations are introduced
\bea
s_0(t)&=& A\sinh \biggl(\dfrac{\pi \Omega t}
         {2{\rm {\bf K'}}}  \biggr),\,\,\,\,
         s_n(t)= (-1)^nA_n\sinh \biggl(\dfrac{(2n+1)\pi \Omega t}
         {2{\rm {\bf K'}}}\biggr),\,\,n=1,...\;. \label{ss},\\\label{A}
         A&=&a\,\frac{\pi}{k
{\rm {\bf {K}}}'}\dfrac{1}{\cosh \biggl( \dfrac{\pi {\rm {\bf K}}}{2{\rm
{\bf K'}}}\biggr)},\,\,\,\,\,A_n=a\,\frac{\pi}{k
{\rm {\bf {K}}}'}\dfrac{1}{\cosh \biggl( \dfrac{(2n+1)\pi {\rm {\bf
K}}}{2{\rm {\bf K'}}}\biggr)} ,\,n=1,... \;. \eea

Let us introduce the notion of $n$-order mode approximations
\bea \label{q-z0}
q_0(t) &=& s_0(t),\\
q_1(t) &=& s_0(t)+s_1(t)\label{q-z1},\\
q_2(t) &=& s_0(t)+s_1(t)+s_2(t)\label{q-z2},\\
...\nonumber\\
q_n(t) &=& s_0(t)+s_1(t)+...\,\,\,s_n(t)\label{q-zn}.\eea

We note that arguments of the $\sinh$ functions and $A_i,\,i=1,...$ are power series of
$k^{\prime 2}$, however in definitions (\ref{q-z0}) - (\ref{q-zn}) we do not make these expansions.

\begin{figure}[h!]
\centering
\includegraphics[width=35mm]{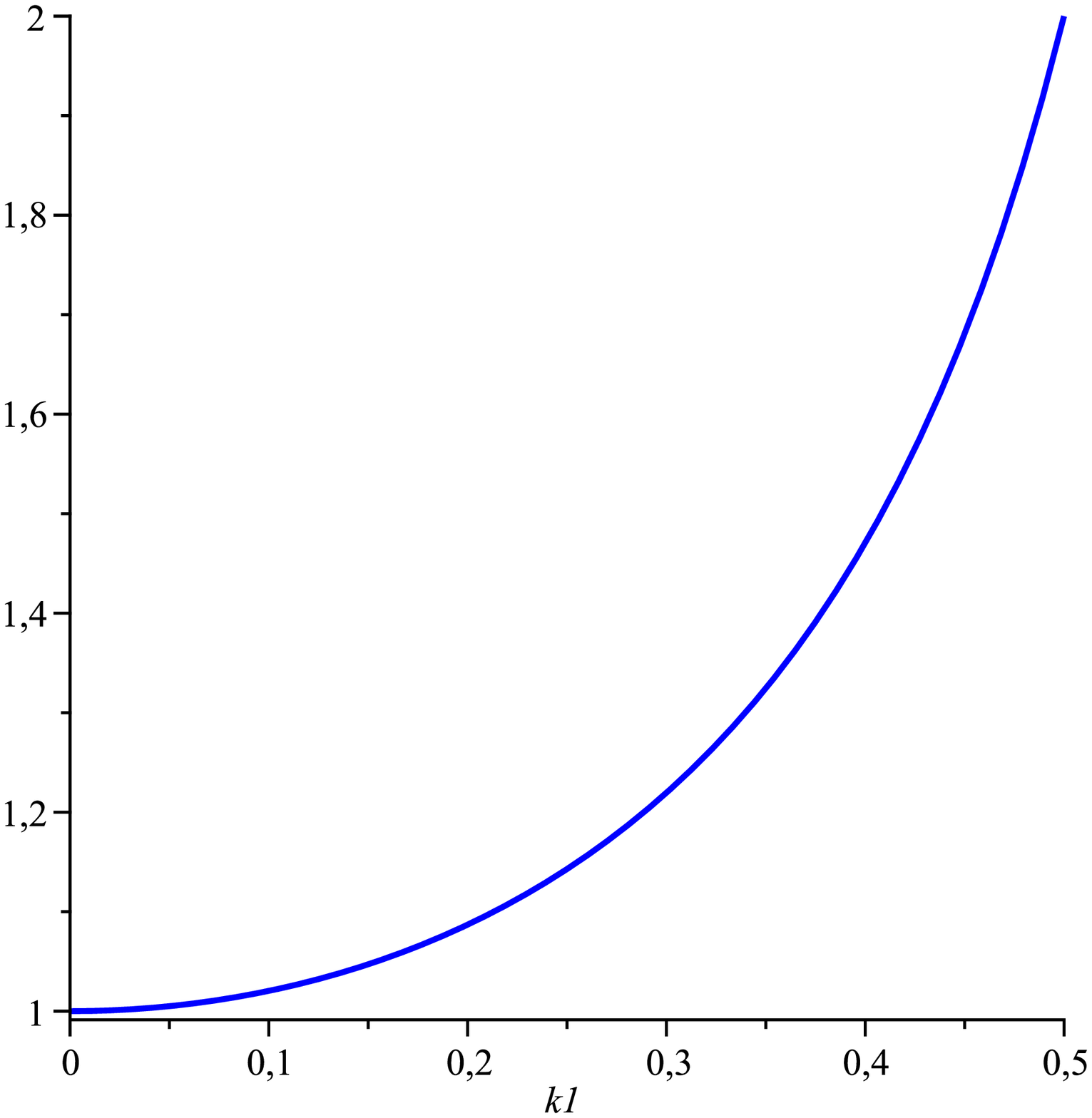}$A.\,\,\,\,\,\,\,\,\,$
\includegraphics[width=35mm]{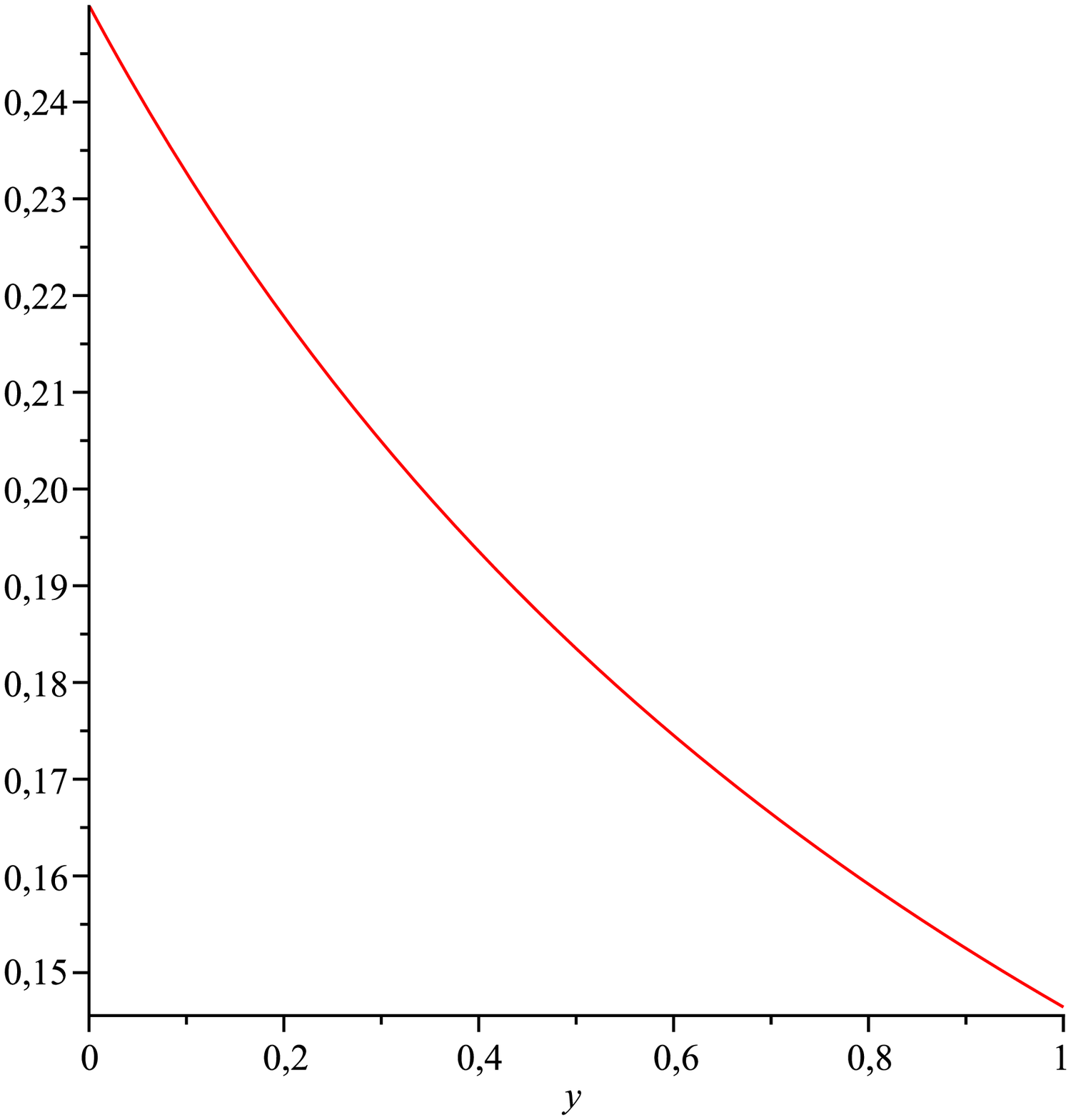}$B.\,\,\,\,\,\,\,\,\,$
\includegraphics[width=35mm]{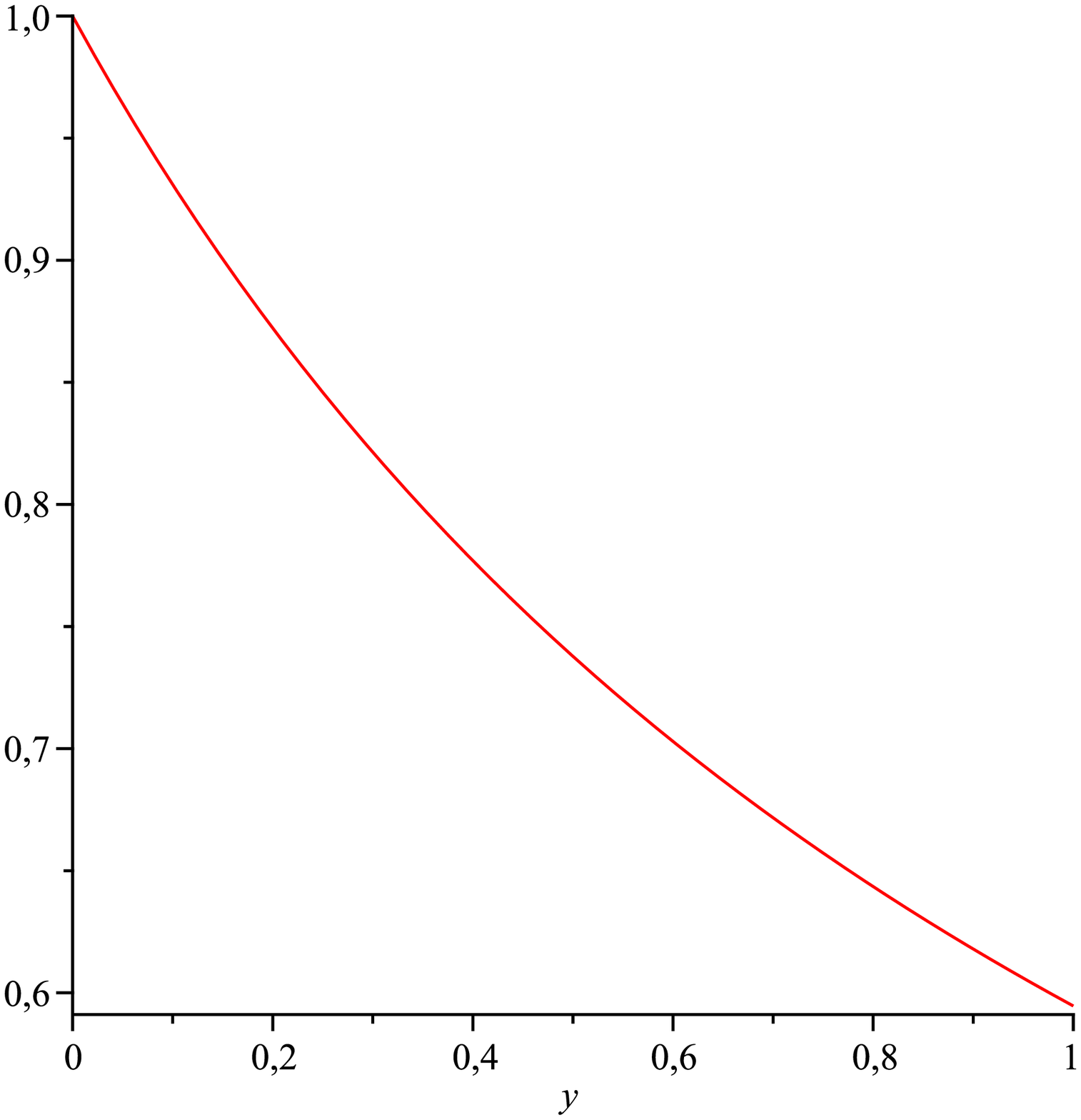}C.
\caption{
A. Estimation (5.12) is shown. 
$k^{\prime 2}$ is plotted along the horizontal axis, and
the left-hand side of (5.12) 
is plotted along the vertical axis.
B. Estimation (5.13) is shown. 
$y$ is plotted along the horizontal axis, and the
left-hand side of (5.13) -  
along the vertical axis. C. Estimation
(5.15) is shown. 
$y$ is plotted along the horizontal
axis, and the left-hand side of
(5.15) 
is plotted along the vertical
axis.} \label{k-ep}
\end{figure}

In order to estimate an error of approximation of (\ref{q-z0}) - (\ref{q-zn}) the following estimations
for $k'<1/2$ are used. The following Lemma provides the estimations.
\begin{lemma}
  For $k'<1/2$ the following estimations hold:
  \begin{enumerate}
    \item
       \be
         \label{ak1m}
          a=\frac{\mu}{\sqrt{\epsilon}}\left(1+\frac{1}{1-2k^{\prime
          2}}\right)^{1/2}< \frac{2\mu}{\sqrt{\epsilon}},
      \ee
    \item
      \be \label{kk-ep}
        k^{'2}=\frac12\biggl(1-\,\frac{1}{\sqrt{1+\frac{4\epsilon E}{\mu^4}}}\biggr)\leq
\frac{2\epsilon E}{\mu^4},
      \ee
    \item
      \be
        \Omega=\mu\,\left(1+\frac{4\epsilon E}{\mu^4}\right)^{1/4}<\mu\,\biggl(1+\frac{4\epsilon E}{\mu^4}\biggr).
      \ee
  \end{enumerate}
\end{lemma}
{\it Proof}   (\ref{ak1m}) follows from the fact that if $k'<1/2$, then (see. Fig. \ref{k-ep}. A)
\be \label{a-k}
\frac{1}{1-2k^{\prime 2}} <2 . \ee
Taking into account (\ref{ak1}) we get (\ref{ak1m}).

The estimation (\ref{kk-ep}) we get in the following way. The estimate holds (see Fig.
\ref{k-ep} B.)
\be \label{eq-k-ep} \frac1{2y}\, \left( 1-{\frac
{1}{\sqrt {1+y}}} \right) <\frac12,\,\,\,\,\,\, 0 \leq y<\infty .\ee

Therefore
\be
\label{kk-ep-m}
  k^{'2}=\frac12(1-\,\frac{1}{\sqrt{1+\frac{4\epsilon E}{\mu^4}}})\leq
\frac{2\epsilon E}{\mu^4}
\ee

The estimate
\be
  \label{Omega-ep} {\frac {(1+y)^{1/4}}{1+\,y}}\leq 1
\ee
is shown in Fig.\ref{k-ep} C. We get
\be
\label{Omega-mu}\Omega=\mu\,\left(1+\frac{4\epsilon
E}{\mu^4}\right)^{1/4}<\mu\,\biggl(1+\frac{4\epsilon E}{\mu^4}\biggr).
\ee


From the estimations (\ref{kk-ep}), (\ref{d1})  and (\ref{ak1m}) the
estimation of the exact solution follows (\ref{q-z0})\bea
\label{q1} \lvert q(t) - q_1(t)\rvert\leq  a d_1\leq 2\mu
\sqrt{\frac{1}{\epsilon}}\dfrac{4}{k }\biggl(\frac{2\epsilon
E}{\mu^4}\biggr)^{3(1-c)/2}= \dfrac{8\sqrt{2E}}{k \mu}\biggl(\frac{2\epsilon
E}{\mu^4}\biggr)^{1-\frac32c}, \eea
for

\be \label{t-rest-m} 0\leq t<\frac{c}{\mu 2\sqrt 2 }\ln \frac{1}{k^{\prime 2}},
\,\,\,\,c<1/2.\ee
Let us note that according to (\ref{kk-ep})
\be
\label{kk-ep-m}
  k^{'2}\leq \frac{2\epsilon E}{\mu^4}
\ee
one has
\be
\label{ep-kk}
  \frac 1{k^{'2}}\geq \frac{\mu^4}{2\epsilon E}\,\,\,\,\Rightarrow\,\,\,\, \ln\frac 1{k^{'2}}\geq
\ln\frac{\mu^4}{2\epsilon E}
\ee
i.e. if the following is true \be
t<\frac{c}{\mu 2\sqrt 2 }\ln\frac{\mu^4}{2\epsilon E}\ee then (\ref{t-rest-m}) holds.

In the same way we get
\begin{lemma}
For $0 < c < 1/2$ then there exists such $\epsilon_0 > 0$ that for
$0 < \epsilon < \epsilon_0$\be \label{rest-t-ep} t<\frac{c}{2\mu \sqrt 2
}\ln\frac{\mu^4}{2\epsilon E}\ee estimations \bea
\label{est-dn-l} |q(t)-q_{n-1}(t)|\leq \dfrac{8\sqrt{2E}}{k
\mu}\left(\frac{2\epsilon E}{\mu^4}\right)^{n-\frac{2n+1}2c}
,\,\,\,n=1,...\eea
hold.
\end{lemma}
{\it  Proof} follows from (\ref{est-dn}) and estimations:
\bea
  \label{est-dn}
  |q(t)-q_{n-1}(t)|\leq \dfrac{4a}{k }
  \biggl(k'\biggr)^{(2n+1)}e^{(2n+1)\Omega t} &<& a\dfrac{4}{k
  }(k')^{(2n+1)(1-c)} \nonumber \\&=& \dfrac{8\sqrt{2E}}{k \mu}\left(\frac{2\epsilon
  E}{\mu^4}\right)^{n-\frac{2n+1}2c}.
\eea

\section{Rearrangement of the series for the elliptic cosine}

By means of the following representation one can obtain the series (\ref{as-t})
 from (\ref{as-tach-cn-m}). Particularly,
the solution (\ref{as-tach-cn-m}) can be presented in the following form:
\begin{equation}\label{tachyon-ser}
q(t) = a\,{\rm {cn}}\,(\Omega t -{\rm {\bf
K}},k)=\sum_{n=0}^{\infty}
(-1)^nA\bar a_n(\lambda)\sinh\left((2n+1)\mu \bar\bO(\lambda)\, t\right) .
\end{equation}
Here
\be\label{lambda-1} \lambda=\frac{A^2\epsilon}{\mu^2}
\ee
and
$\bar a_n(\lambda)$ and $\bar\bO(\lambda)$ are determined by functions
$a_n(k^{\prime 2})$ and $\bO(k^{\prime 2})$:
\bea \label{an}a_n(k^{\prime 2})&=&\frac{\cosh (
\frac{\pi {\rm {\bf K}}}{2{\rm{\bf K'}}})} {\cosh \left(
\frac{(2n+1)\pi {\rm {\bf K}}}{2{\rm {\bf K'}}}\right)},
\,\,\,\,n=1,...,\, \, a_0(\lambda)=1,
\\
\bO(k^{\prime 2})&=& \frac{1}{\mu}\dfrac{\pi \Omega }
         {2{\rm {\bf K'}}}=\frac{1}{\sqrt{1-2k^{\prime 2}}} \dfrac{\pi }
         {2{\rm {\bf K'}}}
         \eea
and via change of variables, i.e.
\bea \label{an-lambda}\bar
a_n(\lambda)&=&a_n\left( F^{-1}(\lambda)\right), \,\,\,\,n=1,...;\,
\, a_0(\lambda)=1,
\\
\label{an-lambda}
\bar\bO(\lambda)&=&\bO\left( F^{-1}(\lambda)\right).
         \eea
$F^{-1}$ is derived from the relation:
\bea\label{A-a-mv} \lambda =F (k^{\prime 2}),\,\,\,\,\,\, F(k^{\prime 2})=
\left(1+\frac{1}{1-2k^{\prime 2}}\right) \left(\dfrac{\frac{\pi}{k
{\rm {\bf {K}}}'}}{\cosh \biggl( \dfrac{\pi {\rm {\bf K}}}{2{\rm {\bf
K'}}}\biggr)}\right)^2  .\eea
One has for the lowest terms o
\be
\lambda=2\, k^{\prime2}+\frac {15}{4}
k^{\prime 4}+{\cal {O}} \left( {k^{\prime 6}} \right). \ee

The proof is obvious. The not so obvious fact is that the series calculated in such a way coincides with the series
(\ref{as-t}),  see Lemma \ref{L3} below.

\subsection{$n$-mode approximation with power series of $k^{\prime}$}
Consider the following power series of $k^{\prime 2}$:

\bea
\label{Omega-k-prime} \bO=\frac{1}{\sqrt{1-2k^{\prime 2}}} \dfrac{\pi }
         {2{\rm {\bf K'}}}&=& 1+k^{\prime 2}\bO _1+k^{\prime 4}\bO _2+...,
          \\
\label{A1-exact-mm} \dfrac{\cosh \biggl( \dfrac{\pi {\rm {\bf K}}}{2{\rm {\bf
K'}}}\biggr)}{\cosh \biggl( (2i+1)\dfrac{\pi {\rm {\bf K}}}{2{\rm {\bf
K'}}}\biggr)}&=&k^{\prime 2i}l_{i\,0}+k^{\prime 2(i+1)}l_{i\,1}+...\;.
\eea
Explicit form of its lowest terms is
\bea
\bO _1=\frac34,\,\,\,\,\,\bO _2=\frac{75}{64},\,\,\,\,l_{1\,0}=\frac1{16},\,\,\,\,\,
l_{1\,1}=\frac{9}{256}.\eea
Using approximation
\bea
  \label{Omega-k-prime-n}\bO^{(n)}&=&(1+k^{\prime 2}
  \bO _1+...k^{\prime 2n}\bO _n),\\
  \label{Ai-appr} A^{(n)}_i&=&A\,\left(k^{\prime 2i}l_{i\,0}+k^{\prime 2(i+1)}l_{i\,1}+...+k^{\prime 2(i+n)}l_{i\,n}\right)
\eea
we will obtain the following approximations for each of the terms in series (\ref{as-tach-cn-m})
\bea
  s^{(0)}_0(t)&=& A\sinh (\mu t)\label{q01},\\
  s^{(1)}_0(t)&=& A\sinh \left((1+k^{\prime 2} \bO _1)\mu t\right)\label{q11},\\
  s^{(n)}_0(t)&=& A\sinh \left((1 +k^{\prime 2} \bO _1+... k^{\prime 2n} \bO
_n)\mu t\right).\label{q21}
\eea
We notice that, we do not expand $A$ in power series of $k^{\prime 2}$.

In the same way,
\bea
s^{(n,j)}_i(t)&=& (-1)^i\, A^{(n)}_i\sinh \left((2i+1)\mu \bO ^{(j)}t\right)\label{r21}, \eea
or explicitly
\be
s^{(n,j)}_i(t)=  (-1)^i\,A\,\left(k^{\prime 2i}l_{i,0}+...+k^{\prime 2(n+i)}l_{i\, n}\right)
\,\sinh \left((2i+1)\mu(1+k^{\prime 2} \bO _1+....k^{\prime j}
\bO _{j})t\right)\label{snij} \ee

In accordance with (\ref{Omega-k-prime}) and (\ref{Ai-appr}) we have the following approximations for $s_1(t)$:
\bea
s^{(0,0)}_1(t)&=& -A\,k^{\prime 2}l_{1,0}\sinh (3\mu t)\label{r01},\\
s^{(0,1)}_1(t)&=& -A\,k^{\prime 2}l_{1,0} \sinh \left(3\mu(1 +k^{\prime 2} \bO _1)t\right)\label{r11},\\
s^{(1,2)}_1(t)&=& -A\,\left(k^{\prime 2}l_{1,0}+k^{\prime 4}l_{1\, 1}\right)\,\sinh \left(3\mu(1 +k^{\prime 2} \bO _1+k^{\prime 4}
\bO _2)t\right)\label{r21}.\eea

We will also use notations:
\bea
  s^{(exact,j)}_i(t)&=&  (-1)^i\,A_i
  \,\sinh \left((2i+1)\mu(1 +k^{\prime 2} \bO _1+...+k^{\prime 2j}
  \bO _j)\,t\right)\label{sni},\\
  s^{(n,exact)}_i(t)&=&  (-1)^i\,A\,\left(k^{\prime 2i}l_{i,0}+...+k^{\prime 2(n+i)}l_{i\, n}\right)
  \,\sinh \left((2i+1)\mu\bO\, t\right)\label{sni}.
\eea

\subsection{$n$-mode approximation with power series of $\lambda$}
In order to express the series (\ref{as-tach-cn-m}) in a way analogous to
(\ref{as-t}) we have to expand $\bO$ and $A_i/A$ in power series of $\epsilon$
 \bea
\label{Omega-appr-e}\bO&=&\mu\,\left[\,1 +\frac{A^2\epsilon}{\mu^2}
\bar\bO _1+\left(\frac{A^2\epsilon}{\mu^2}\right)^2 \bar\bO _2+...\right] ,\\
\label{Ai-appr-e} A_i&=&A\,\left[\left(\frac{A^2\epsilon}{\mu^2}\right)^i\bar l_{i0}
+\left(\frac{A^2\epsilon}{\mu^2}\right)
^{i+1}\bar l_{i\,1}+...\right] ,\\\label{bar-Omega1}
\bar\bO _1&=&\frac38,\,\,\,\,\,\bar l
_1=-\frac{1}{32}.\eea
We will also use notations:
\bea
\label{Omega-appr-e}\bO^{(n)}&=&\mu\,\left[\,1 +\frac{A^2\epsilon}{\mu^2}
\bar\bO _1+\left(\frac{A^2\epsilon}{\mu^2}\right)^2 \bar\bO _2+...+\left(\frac{A^2\epsilon}{\mu^2}\right)^n \bar\bO_n\right] ,\\
\label{Ai-appr-en} A^{(n)}_i&=&A\,\left[\left(\frac{A^2\epsilon}{\mu^2}\right)^i\bar l_i+\left(\frac{A^2\epsilon}{\mu^2}\right)
^{i+1}\bar l_{i\,1}+...+\left(\frac{A^2\epsilon}{\mu^2}\right)^{i+n}\bar l_{i\,n}\right] .
\eea
In the lowest terms we have the following approximations
\bea
\bar s^{(0)}_1(t)&=&A\,
\sinh \left(\,\mu t\right)\label{r0-bar} ,\\
\bar s^{(1)}_0(t)&=& A \,
\sinh \left((\,1 +\frac{A^2\epsilon}{\mu^2}
\bar\bO _1)\,\mu t\right)\label{r01-bar} ,\\
\bar s^{(0,1)}_1(t)&=& -\frac{A^3\epsilon}{\mu^2}\,\bar l_1 \,
\sinh \left(3(\,1 +\frac{A^2\epsilon}{\mu^2}
\bar\bO _1)\,\mu t\right)\label{r11-bar} ,\\
\bar s^{(1,2)}_1(t)&=& -A\, \left(\,\frac{A^2\epsilon}{\mu^2}\,\bar
l_1 + \left(\frac{A^2\epsilon}{\mu^2}\right)^2\bar l_2\right)\, \nonumber\\&\sinh&
\left(\,3\,\left(\,1 +\frac{A^2\epsilon}{\mu^2} \bar\bO
_1+\left(\frac{A^2\epsilon}{\mu^2}\right)^2 \bar\bO _2\right)\,\mu
t\,\right)\label{r21-bar}
\eea
here $\bar\bO_1$ and $\bar l_1$ are given by (\ref{bar-Omega1}). The bar over $s_i$ stays for expansions in power series of $\epsilon$ (or $\lambda$).

We have the following Lemma.
\begin{lemma}\label{L3}
  When amplitudes (\ref{A}) and  (\ref{as-t-hyp}) are equaled, i.e.
  \be
    \mA=A
  \ee
  the following relations hold
  \bea
    \ma_{n\, k}&=&\bar l_{n\,k},\\
    \mm\, _{i}&=&\bar \bO _{i}.
  \eea
\end{lemma}

Lemma (\ref{L3}) allows to claim that if $\mA=A$ then
 \bea
\ms^{(i)}_0(t)&=& \bar s^{(i)}_0(t),\,\,\,i=1,....\,\, , \\
\ms^{(i,j)}_n(t)&=& \bar s^{(i,j)}_n(t),\,\,\,\,\,n=1....,i,j=0,1...\,.\eea

\section{Hyperbolic  analogue of the modified perturbation theory}

The goal of the discussions below is to obtain estimation of error of approximation in case
 (\ref{as-t-hyp}) (mentioned at the beginning of the paper) is used. To do this we note that the constant $A_i$
in (\ref{q-zn}) is  a power series of $k^{\prime 2}$ (or a power series of $\epsilon$).
Also the arguments of the sinh functions and the $\sinh$ functions are power series
of $k^{\prime 2}$ and
consequently the the power series of $\epsilon$.

\subsection{One mode approximations}
Approximations to the exact one mode expression
\be
\label{q0}
q_0(t)=A\sinh \left(\left(
 \dfrac{\pi \Omega }
         {2{\rm {\bf K'}}}  \right) t\right)\ee
have the form
\bea\label{s0}
\bar s^{(0)}_0(t)&=&A\sinh (\mu t), \\
\bar s^{(1)}_0(t)&=&A\sinh \biggl(\mu\biggl(1+\frac{3A^2}{8\mu^2}\epsilon t\biggr)\biggr)
\label{s1}, \\
\label{s2}
\bar s^{(2)}_0(t)&=&A\sinh \biggl(\mu\biggl(1+\frac{3A^2}{8\mu^2}\epsilon - \dfrac{15 A^4}{2^8 \mu^4}\epsilon^2 \biggr)t\biggr). \eea

In relations (\ref{s0}) - (\ref{s2}) the $\sinh$-functions arguments are different from ones in (\ref{q0}). We denote these differences by $\Delta_n$: 
\bea
\Delta_0&=&\left(\mu-
 \dfrac{\pi \Omega }
         {2{\rm {\bf K'}}}  \right)t, \\
         \Delta_1&=&\left(\mu\biggl(1+\frac{3A^2}{8\mu^2}\epsilon \biggr)-
 \dfrac{\pi \Omega }
         {2{\rm {\bf K'}}}  \right)t, \\
         \Delta_2&=&\left(\mu\biggl(1+\frac{3A^2}{8\mu^2}\epsilon- \dfrac{15 A^4}{2^8 \mu^4}\epsilon^2 \biggr)-
 \dfrac{\pi \Omega }
         {2{\rm {\bf K'}}}  \right)t.
         \eea

Let us estimate contribution due to the mentioned differences. We have
\bea \label{q-q}
s_0(t)-s ^{(n)}_0(t)&=&A\left[\sinh\left(\mu\bO\, t\right)-
\sinh\left(\mu \bO\, t+\Delta^{(n)}\right)\right]\\& =&
A\left[\sinh\left(\mu \bO \, t\right)-
\sinh\left(\mu \bO\,  t\right)\cosh(\Delta_n)\right.\nonumber
-\left.\cosh\left(\mu \bO \, t\right)\cdot\sinh(\Delta_n)\right]\\
&=& A\sinh\left(\mu \bO\, t\right)(1-\cosh(\Delta_n))- A\cosh\left(\mu \bO\,
t\right)\cdot\sinh(\Delta_n) \nonumber\eea
Taking into account that for small $\Delta$ (for instance, $0<\Delta<1$)
the following relations hold\bea \label{sinh-Delta-m}
\sinh(\Delta)&< &2 \Delta,\\
\label{cosh-Delta-m} |1-\cosh(\Delta)|&< &2 \Delta.\eea
Thus we get for $t>0$ and
\begin{equation}
  \label{delta-1}
  \Delta_n < 1,
\end{equation}
\bea
  \nonumber
  |s_0(t)-s ^{(n)}_0(t)|&<& 2\Delta_n A\sinh\left(\mu \bO  t\right)+
  2\Delta_n A\cosh\left(\mu \bO  t\right)\\
  \label{q-q1}&=& 2\Delta_n Ae^{\mu \bO  t}.\\
\eea

At this point one can see that an error of the approximation of $s_0(t)$ by $s^{(n)}_0(t)$ is defined by infinitesimality of $\Delta_n$.

In particular, for $\Delta_0$ we have
\bea
\Delta_{0}&=&\mu t\delta_0,
\\
\delta_0&=&1-
 \dfrac{\pi }
         {2{\rm {\bf K'}}} \frac{1 }{\sqrt{1-2k^{\prime 2}}}.\label{delta-2mmm}
         \eea
\begin{figure}[h!]
\centering
\includegraphics[width=45mm]{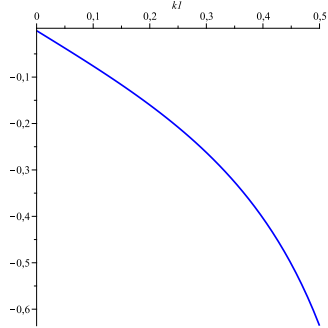}\label{mu-omega-m}
\caption{
The estimate (7.15) 
is showed. Along the horizontal axis $k^{\prime2}$ are plotted, and
on vertical axis the relation the relation of the right-hand side of
(7.15) 
to $k^{\prime 2}$.}\label{mu-omega-m}
\end{figure}

From analysis  (see Fig. \ref{mu-omega-m}) we get
       \be
       \label{mu-Omega}
      \delta _0<\sqrt{2}k^{\prime 2}, \,\,\,\,\,\,\,0\leq k^{\prime }\leq \frac12
      \ee
      and therefore,
      \be
        \label{mu-Omega-m}
        \Delta_0<\sqrt{2}\mu k^{\prime 2}t, \,\,\,\,\,\,\,0\leq k^{\prime }\leq \frac12.
      \ee

Estimates $\Delta_n$, $n=1,2$ are presented in Appendix.B.
\subsection{Two modes approximations}
\begin{figure}[h!]
\centering
\includegraphics[width=45mm]{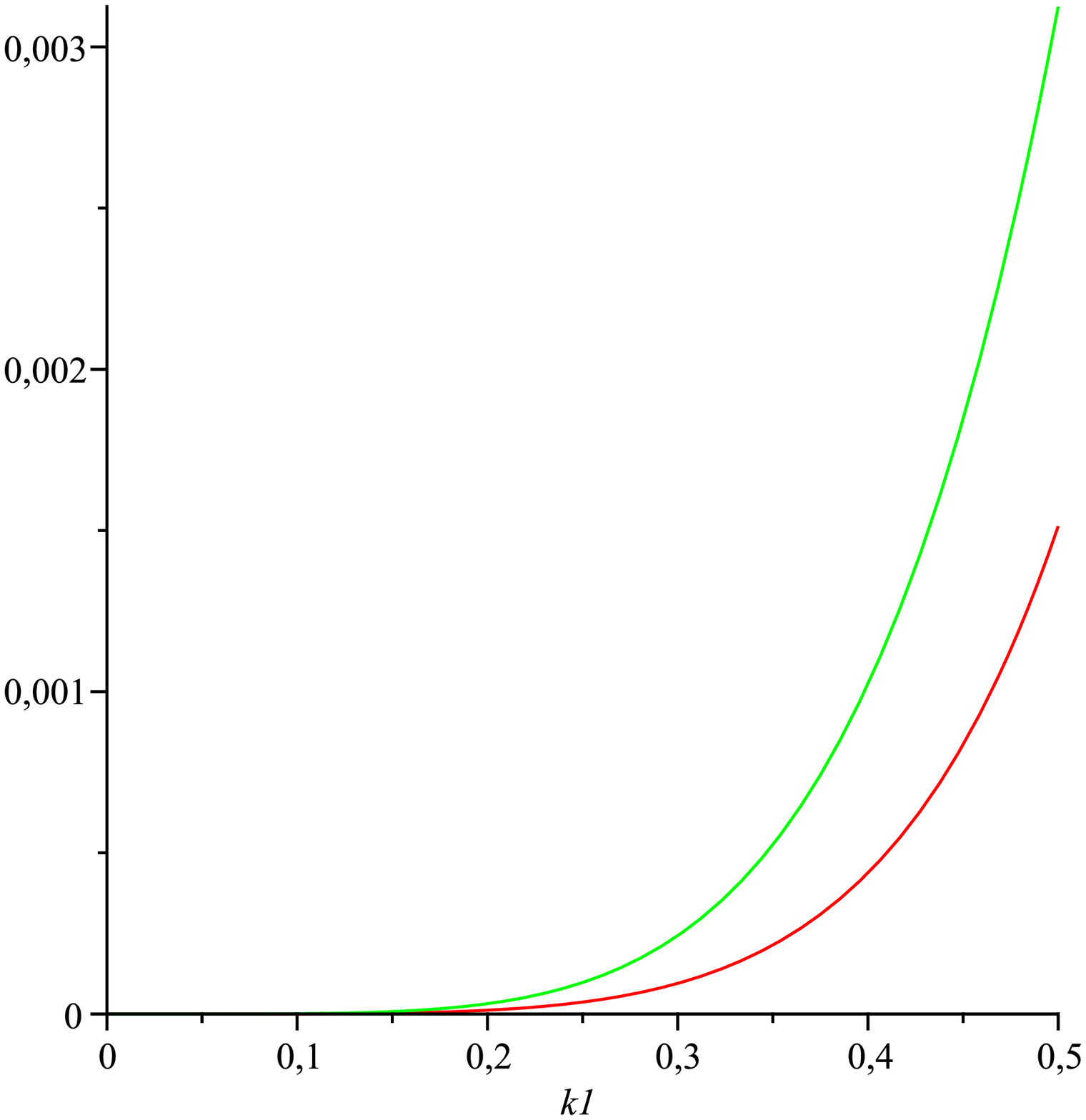}$A\,\,\,\,\,\,\,\,\,\,$
\includegraphics[width=45mm]{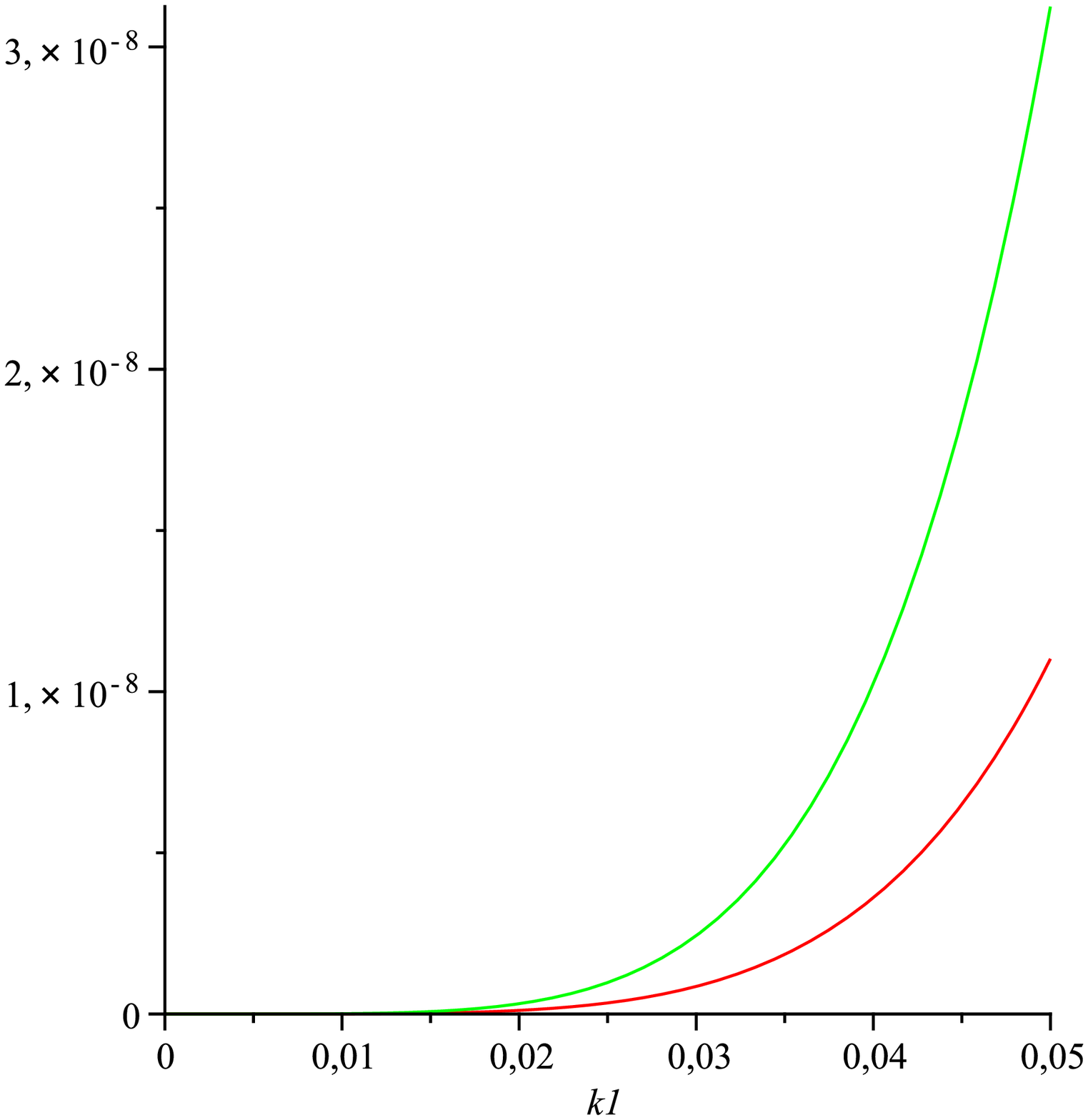}$B\,\,\,\,\,\,\,\,\,\,$
\caption{(color online)
A. The estimation
(7.25) 
is illustrated. On the horizontal axis $k^{\prime 2}$
is plotted, on the vertical axis the values of  the left-hand side of
(7.25) 
(red color) and the right-hand side (7.25) 
(green color) are presented. B.
The zoom of the plot in the left panel A for small values of $k^{\prime 2}$.} \label{A1-estf}
\end{figure}

Consider the two modes term $q_1(t)$, \bea
q_1(t) &=& q_0(t)+s_1(t),\eea
where
\bea
s_1(t)&=& -A_1\sinh \biggl(\dfrac{3\pi \Omega t}
         {2{\rm {\bf K'}}}  \biggr),\,\,\,\,\,\,\,n=1,...,\\
         A_1&=&a\,\frac{\pi}{k
{\rm {\bf {K}}}'}\dfrac{1}{\cosh \biggl( \dfrac{3\pi {\rm {\bf K}}}{2{\rm
{\bf K'}}}\biggr)}
\eea
$s_1(t)$ is approximated by
\be
\bar s^{(0,0)} _1=- \frac{\epsilon A^3}{32\mu^2}\sinh\left(3\mu t\right)
\ee
or
\be s^{(0,0)}_1(t)=- A\,k^{\prime 2}\frac{1}{16}\sinh (3\mu t). \label{r11m}\ee
A deviation of $\bar s^{(0,0)} _1$ from the exact $s_1(t)$ is not only due to
changes in the argument of  the $\sinh$-function, $3 \Delta _0$, but also due to corrections to the exact value of $A_1$:
\be A_1= \frac1{16}
k^{\prime 2}A+a\,k^{\prime 5}L_{1}(k')\ee and we have an estimation
\be
  L_{1}<0.1\,\,\,\,for \,\,\,\,\,k'<1/2.
\ee
This estimation is true since (see Fig. \ref{A1-estf})
 \be\label{A1-est}
-\frac{k^{\prime 2}}{16}\frac{\pi}{k
{\rm {\bf {K}}}'}\dfrac{1}{\cosh \biggl( \dfrac{\pi {\rm {\bf K}}}{2{\rm
{\bf K'}}}\biggr)}+\frac{\pi}{k
{\rm {\bf {K}}}'}\dfrac{1}{\cosh \biggl( \dfrac{3\pi {\rm {\bf K}}}{2{\rm
{\bf K'}}}\biggr)}<\frac1{10}\,k^{\prime 5}.\ee

We  get
\bea
s^{(0,0)} _1(t)-s _1(t)&=&- A\,k^{\prime 2}\frac{1}{16}\sinh (3\mu t)+
A_1\sinh (3\mu\bO t )\nonumber\\
  &=&- A\,k^{\prime 2}\frac{1}{16}\sinh (3\mu (\delta _0 +\bO)t)+
A_1\sinh (3 \mu \bO t)\nonumber\\
 &=&- A\,k^{\prime 2}\frac{1}{16}\left(\sinh (3\mu (\delta _0+\bO)t)-
\sinh (3\mu  \bO t ) \right)\nonumber \\&+&a(- \,\frac{k^{\prime 2}}{16}A+A_1)
\sinh (3\mu \bO t).\eea
Consequently,
\bea
|s^{(0,0)} _1(t)-s _1(t)| &<&- A\,k^{\prime 2}\frac{1}{16}
\left|\,\sinh (3\mu (\delta _0 -\bO)t)-
\sinh (3\mu \bO t) \right|\\
&+& a\frac {k^{\prime 5}}{10}  |\sinh (3\mu \bO )t|.   \eea

Taking into account an estimation of the form (\ref{q-q1}) we get
\be
\label{s001}
|s^{(0,0)} _1(t)-s _1(t)|<Ck^{\prime 4}\,e^{3\mu \bO t}, \ee
where $C$ is a constant.

\subsection{The second order  of
 the modified perturbation theory approximation}

The following theorem holds
\begin{theor}
For
\be
\label{cond-T1}
0<t<\frac{c}{2\sqrt2\,\mu
}\ln\frac{\mu^4}{2\epsilon E}\,\,\,\,\,
\mbox{{\rm and}}
\,\,\,\,\,0<c<\frac12
\ee
there exist such $C$ and $\sigma<3/4$  that, the following estimation  is true
\bea
\label{est-m-2} |q(t)-s_0^{(1)}(t)-s_{1}^{(0,0)}|\leq C\dfrac{\sqrt{E}}{
\mu}\left(\frac{\epsilon E}{\mu^4}\right)^{2-\sigma}
,\,\,\,n=1,...\eea
\end{theor}

{\it Remark.} In the inequality  (\ref{est-m-2}) $C$ -- dimensionless constant,
factor $\dfrac{\sqrt{E}}{\mu}$ is introduced to get   dimensionless quantity,
$\frac{\epsilon E}{\mu^4}$ -- dimensionless factor.

{\it  Proof.} The proof follows from the estimates (\ref{est-dn-l})
and   (\ref{s001}). Indeed, from (\ref{est-dn-l}) it follows that
\bea
\label{est-dn} |q(t)-q_1(t)|\leq \dfrac{8\sqrt{2E}}{k
\mu}\left(\frac{2\epsilon E}{\mu^4}\right)^{2-\frac{3}2c}
,\,\,\,n=1,...\; .\eea
In accordance with (\ref{q-z1})
\be
  q_1(t) = q_0(t)+s_1(t)\label{q-z1-m}
\ee
and we have
\be
q(t)-s_0^{(1)}(t)-s_{1}^{(0,0)}=q(t)-q_1(t)+q_0(t)-s_0^{(1)}(t)+s_1(t)-s_{1}^{(0,0)}(t),\ee
and consequently
\be
|q(t)-s_0^{(1)}(t)-s_{1}^{(0,0)}|<|q(t)-q_1(t)|+|q_0(t)-s_0^{(1)}(t)|+|s_1(t)-s_{1}^{(0,0)}(t)|\ee

In order to finish the proof we need the estimates (\ref{delta-1})
and (\ref{delta-4m}) for (\ref{delta-4}). Taking into account these estimations we get
(\ref{est-m-2}).

\subsection{Estimation of the  $n$-mode approximation by power series of $k^{\prime 2}$}

Consider the $n$th mode $s_n(t)$,
\bea
s_n(t)&=& (-1)^nA_n\sinh \biggl(\dfrac{(2n+1)\pi \Omega t}
         {2{\rm {\bf K'}}}\biggr)\nonumber\\&=&(-1)^nA\, l_{n}\sinh \left({(2n+1)\mu \bO t}\right),\,\,\,n=1,...,\eea
         where\bea
         A_n&=&a\,\frac{\pi}{k
{\rm {\bf {K}}}'}\dfrac{1}{\cosh     \biggl( \dfrac{(2n+1)\pi {\rm {\bf K}}}{2{\rm
{\bf K'}}}\biggr)} \nonumber\\&=& A\dfrac{\cosh \biggl( \dfrac{\pi {\rm {\bf K}}}{2{\rm
{\bf K'}}}\biggr)}{\cosh \biggl( \dfrac{(2n+1)\pi {\rm {\bf K}}}{2{\rm
{\bf K'}}}\biggr)} =\dfrac{A\cosh ( {\rm {\bf K}}\bO )}{\cosh ( (2n+1){\rm {\bf K}} \bO)}\eea
and
\be
  \label{l-n}
  l_n(k')=\dfrac{\cosh ( {\rm {\bf K}}\bO )}{\cosh ( (2n+1){\rm {\bf K}} \bO)}.
\ee

As an approximation to $s_n(t)$ we take
\bea
 s^{(r,m)} _n&=&(-1)^n A\, l^{(r)}_{n}\sinh \left({(2n+1)\mu \bO ^{(m)}t}\right)\eea
where the series $ l^{(r)}_{n}$
is obtained from the expansion of the right-hand side of (\ref{l-n})
\bea l^{(r)}_{n}&=&l_{n,0}k^{\prime 2n}+...+l_{n,r}k^{\prime 2(n+r)}
\eea
and $\bO ^{(m)}$ is defined in (\ref{Omega-k-prime})

 \bea\label{bar_Omega_kprime}\bO ^{(m)}&=&1+\bO _{1}k^{\prime 2}+...+\bO _{m}k^{\prime 2m}\eea

The following lemmas are true \cite{AV}.

\begin{lemma}\label{lemma-n}
For $0<k'<1/2,$ $r \geq 1, n \geq 1,$ there exist constants $B^{(r)}$ and $L_n^{(r)}$ such that
\bea
|\bO(k')-\bO^{(r)}(k')|&<&B^{(r)}k^{\prime 2(r+1)}\\
|l_{n}(k')-l^{(r)}_{n}(k')|&<&L_n^{(r)}k^{\prime 2(n+r+1)}.\eea
\end{lemma}

From Lemma \ref{lemma-n} one gets the following
\begin{lemma}\label{lemma-sn}
\bea
 |s_n(t)-s^{(r,m)} _n(t)|&=&S^{(r,m)} _nk^{\prime 2(n+r+1)}e^{\sigma^{(r,m)}_n t}
 \eea\end{lemma}

$$\,$$
\subsection{ Estimation of the  $n$-mode approximation by power series of $\lambda$}

The $n$-mode can be considered as a function of $\lambda$:
 \bea
s_n(t)&=&(-1)^nA\, \bar l_{n}(\lambda)\sinh \left({(2n+1)\mu \bar \bO(\lambda) t}
\right),\,\,\,\,\,\,\,\,\,\,\,\,\,\,\,\,\,n=1,...\eea
Using the power series of $\lambda$ one introduces notations
\bea
 \bar s^{(r,m)} _n&=& A\, \bar l^{(r)}_{n}\sinh \left({(2n+1)\mu \bar\bO ^{(m)}t}\right)\eea
where $\bar l^{(r)}_{n}$ and $\bar\bO ^{(m)}$ are obtained
 expanding of the right-hand side of (\ref{l-n}) and (\ref{bar_Omega_kprime}) in the power series of $\lambda$
\bea l^{(r)}_{n}&=&\bar l_{n,0}\lambda^{n}+...+\bar l_{n,r}\lambda^{(n+r)},
\eea
\bea\bar\bO ^{(m)}&=&1+\bar\bO _{1}\lambda^{ 2}+...+\bar\bO _{m}\lambda^{m}\eea

The following lemmas are true.

\begin{lemma}\label{lemma-n-lambda}
For $\lambda$ small enough there exist constant $\bar B^{(r)}$ and $\bar L_n^{(r)}$ such that
\bea
|\bar\bO(\lambda)-\bO^{(r)}(\lambda)|&<&\bar B^{(r)}\lambda^{(r+1)}\\
|\bar l_{n}(\lambda)-\bar l^{(r)}_{n}(\lambda)|&<&\bar L_n^{(r)}\lambda^{(n+r+1)}.\eea
\end{lemma}

From Lemma \ref{lemma-n} it follows that
\begin{lemma}\label{lemma-sn}
\bea
 |\bar s_n(t)-\bar s^{(r,m)} _n(t)|&=&\bar S^{(r,m)} _n\lambda^{(n+r+1)}e^{\bar \sigma^{(r,m)}_n t}
 \eea\end{lemma}

$$\,$$
The following theorem holds \cite{AV}:
\begin{theor}
For
\be
  \label{cond-T1} 0<t<\frac{c}{2\sqrt2\,\mu}\ln\frac{\mu^4}{2\epsilon E}\,\,\,\,\,\mbox{{\rm and}}\,\,\,\,\,0<c<\frac12
\ee
one can find such constants $C_n$, $\sigma<3n/4$ and $\epsilon_0 > 0$ that, the following estimate is true for all $0 < \epsilon < \epsilon_0$
\bea
  \label{est-m-n} |q(t)-q^{(n)}(t)|\leq C_n\dfrac{\sqrt{E}}{\mu}\left(\frac{\epsilon E}{\mu^4}\right)^{2n-\sigma_n},\,\,\,n=1,...,\eea
  where
\be
  \label{est-n-m}
  q^{(n)}(t)=\sum _{i=0}^n\bar s^{(n-i,n-i)} _i(t).
\ee
In the right-hand side of (\ref{est-n-m}) contributions to all of the first $n$-modes
are taken with orders not exceeding $(n-i)$.
\end{theor}

{\it Remark.} $C_n$ is a dimensionless constant, the factor
$\dfrac{\sqrt{E}}{\mu}$ is introduced by the dimension analysis,
$\frac{\epsilon E}{\mu^4}$ is a dimensionless factor, $\sigma_n$
depends on $n$.

{\it Proof} follows from the estimates (\ref{est-dn-l}) and (\ref{s001}).
Indeed, from (\ref{est-dn-l}) it follows that
\bea
\label{est-dn} |q(t)-q_n(t)|\leq \dfrac{8\sqrt{2E}}{k
\mu}\left(\frac{2\epsilon E}{\mu^4}\right)^{2-\frac{2n+1}2c}
,\,\,\,n=1,...\;.\eea
According to (\ref{q-zn})
\be
q_n(t) = s_0(t)+s_1(t)+...\,\,\,+\,s_n(t)\label{q-zn-m}\ee
and in (\ref{est-n-m}) we take the approximation of $s_0^{(n)}(t)$ to $q_0(t)$ and the approximation
$s_{1}^{(n-1,n-1)}$, for $s_1(t)$ and so on. We get
\bea
q(t)-q^{(n)}(t)&=&q(t)-s_{0}(t)-...-s_{n}(t)\\&+&s_{0}(t)-\bar s_0^{(n)}(t)+
s_{1}(t)-\bar s_{1}^{(n-1,n-1)}(t)+....
+s_n(t)-\bar s_{n}^{(0,0)}(t)
\nonumber\eea
and consequently
\bea
|q(t)-q^{(n)}(t)|&<&|q(t)-q_n(t)|+|q_0(t)-\bar s_0^{(n)}(t)|\nonumber\\ &+& |s_1(t)-\bar s_{1}^{(n-1,n-1)}(t)| + \,...\,+|s_n(t)-\bar s_{n}^{(0,0)}(t)|.\eea

In order to finish the proof we need estimates (\ref{delta-1}) and the estimate (\ref{delta-4m}) for (\ref{delta-4}). Having taken into account these relations we get (\ref{est-m-n}).

\section{Conclusion}
The method for approximate solution to nonlinear dynamics equations
in the  rolling    regime is presented. It is shown that in order to
improve perturbation theory in the rolling   regime it turns out to be
effective not to use an expansion in trigonometric functions as it
is done in case of small oscillations but use expansions in
hyperbolic functions instead. In particular, the Higgs equation in
the rolling    regime is considered. This     regime is investigated using
the representation of the solution  in terms of elliptic functions.
An accuracy of the corresponding approximation is estimated.

In this paper we have investigated the rolling    regime
for motions starting from the top
of the Higgs potential, see Fig.\ref{pot}.A. The same method can be used for motion starting with zero velocity
from the side that is near to the top, see Fig.\ref{pot}.C.

As to possible cosmological applications of rolling solutions it is
worth to  mention that in cosmology various rolling solutions are widely used.
There is a notion of the slow
roll    regime that means that one can ignore the second order
derivatives in the equation of motion \cite{Mbook,Rubakov}.
The rolling from the top of the potential is not
necessary described by the slow roll approximation  and methods
similar to ones developed in this paper can be applied \cite{ABG}.

There are rolling solutions also considered
in the nonlocal cosmology \cite{IA04}-\cite{Barnaby:2008vs}. These
solutions have their  analogue in the flat space-time \cite{YaV-VV,Volovich:2003,AJK}.
Non slow roll regime can be important in inflation and it is related
to so-called stretch effect in nonlocal theories
\cite{Barnaby:2006,Lidsey:2007,BarCline}.

\section*{Acknowledgments}
The present work is partially supported by the following grants:
RFFI 11-01-00894-a(I.A.), NSch-4612.2012.1.(I.A.),
 RFBR 11-01-00828-a (E.P. and I.V), NSch-2928.2012.1 (E.P. and I.V).

\appendix

\section{Proof of Lemma 1 }
Consider first in the case $n=1$.
The difference in the left-hand side of (\ref{estim-m}) is
\begin{eqnarray}
\Delta _1\equiv{\rm {cn}}(\Omega t-{\rm {\bf {K}}},k) -\dfrac{\pi}{k
{\rm {\bf {K}}}'}\dfrac{\sinh \biggl(\dfrac{\pi \Omega t}{2{\rm {\bf K'}}}
\biggr)}{\cosh \biggl( \dfrac{\pi {\rm {\bf K}}}{2{\rm {\bf K'}}}\biggr) }. \eea
Using (\ref{as-tach-cn}) and the majorant of the following sum
 \bea
 r_1\equiv\dfrac{\pi}{k {\rm {\bf
{K}}}'} \left\{-\dfrac{\sinh (3\rho' u'')}{\cosh ( 3\rho ')}+
\,\dfrac{\sinh ( 5\rho' u'')}{\cosh ( 5\rho ')}+...\right\}, \eea
where \be u'' = \dfrac{\Omega t}{{\rm {\bf
K}}},\,\,\,\,\,\,\,\,\,\,\,\,\,\,\,\,\rho' = \dfrac{\pi {\rm {\bf
K}}}{2{\rm {\bf K'}}} ,\,\,\,\,\,\,\,\,\,\,\,\,\,\,\,\,u''\rho'=u' =
\dfrac{\pi\Omega t}{2{\rm {\bf K}'}}
\ee
we get
\bea |r_1| \leq
\dfrac{\pi}{k {\rm {\bf {K}}}'} \sum\limits_{j =
1}^{\infty}\dfrac{\sinh ( (2j+1)\rho' u'')}{\cosh ( (2j+1)\rho ')}
<\dfrac{\pi}{k {\rm {\bf {K}}}'} e^{-\rho'(1 - u'')} \sum\limits_{j
= 1}^{\infty}e^{-2j\rho'(1 - u'')}. \nonumber \eea
Using
\bea
\dfrac{\pi}{k {\rm {\bf {K}}}'} e^{-\rho'(1 - u'')} \sum\limits_{j =
1}^{\infty}e^{-2j\rho'(1 - u'')}= \dfrac{\pi}{k {\rm {\bf {K}}}'}
\dfrac{e^{-3\rho'(1 - u'')}} {1 - e^{-2\rho'(1 -
u'')}},\label{prog-m} \eea
we get
\be
|r _1|\leq d_1,\label{main_estm-m}
\ee
where
\be d_1\equiv\dfrac{\pi}{k {\rm
{\bf {K}}}'} \dfrac{e^{-3\rho'(1 - u'')}} {1 - e^{-2\rho'(1 -
u'')}}. \label{main-d} \ee

In Fig.\ref{F:estimation-delta2-m}. A. the bound (\ref{main_estm-m}) is shown.

We notice that the series in (\ref{prog-m}) converges if
\bea
  e^{-\rho'(1 - u'')}<1,
\eea
that is true if \be 0 < u'' < 1\label{u''-m} \ee
since
 \bea
 \rho' = \dfrac{\pi {\rm {\bf K}}}{2{\rm {\bf K'}}}>0
\eea
Since $u''  = \dfrac{\Omega t}{{\rm {\bf K}}}$ the relation
(\ref{u''-m}) means that \be 0<\dfrac{\Omega t}{{\rm {\bf K}}}<1.\ee

Consider the right-hand side of (\ref{main_estm-m})
 \bea
d_1\equiv\dfrac{\pi}{k {\rm {\bf {K}}}'} \dfrac{e^{-3\rho'(1 -
u')}}{1 - e^{-2\rho'(1 - u')}} =\dfrac{\pi}{k {\rm {\bf {K}}}'}
\dfrac{\exp\left(-3\dfrac{\pi {\rm {\bf {K}}}}{2{\rm {\bf
{K'}}}}\right)\exp\left(\dfrac{3\pi\Omega t}{2{\rm {\bf
{K'}}}}\right)}{1 - \exp\left(-2\dfrac{\pi {\rm {\bf {K}}}}{2{\rm
{\bf {K'}}}}\right)\exp\left(\dfrac{2\pi\Omega t}{2{\rm {\bf
{K'}}}}\right)}. \label{RHS-m}\eea

\begin{figure}[h!]
\centering
\includegraphics[width=35mm]{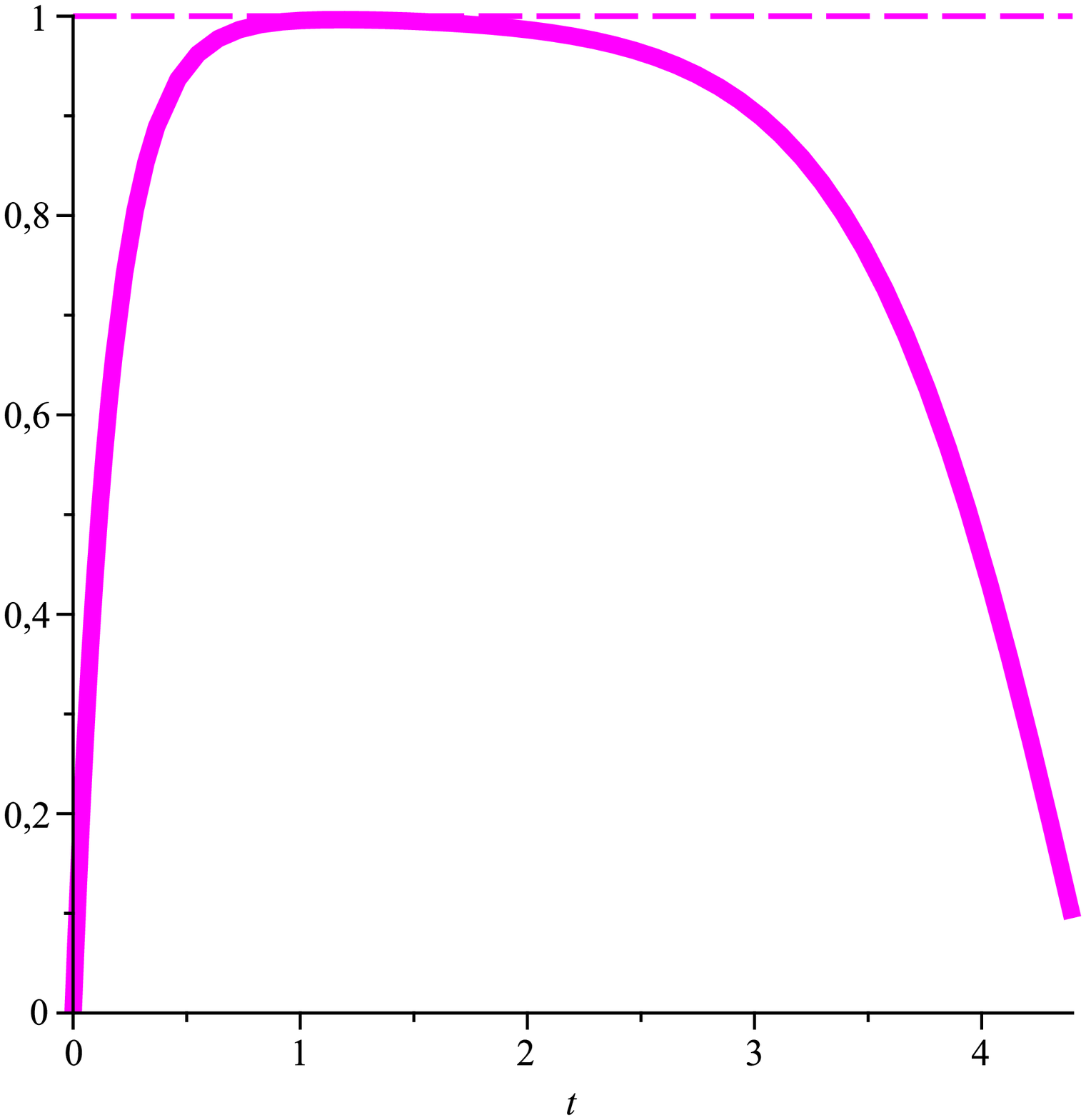}A.$\,\,\,\,\,\,\,\,\,\,\,\,$
\includegraphics[width=35mm]{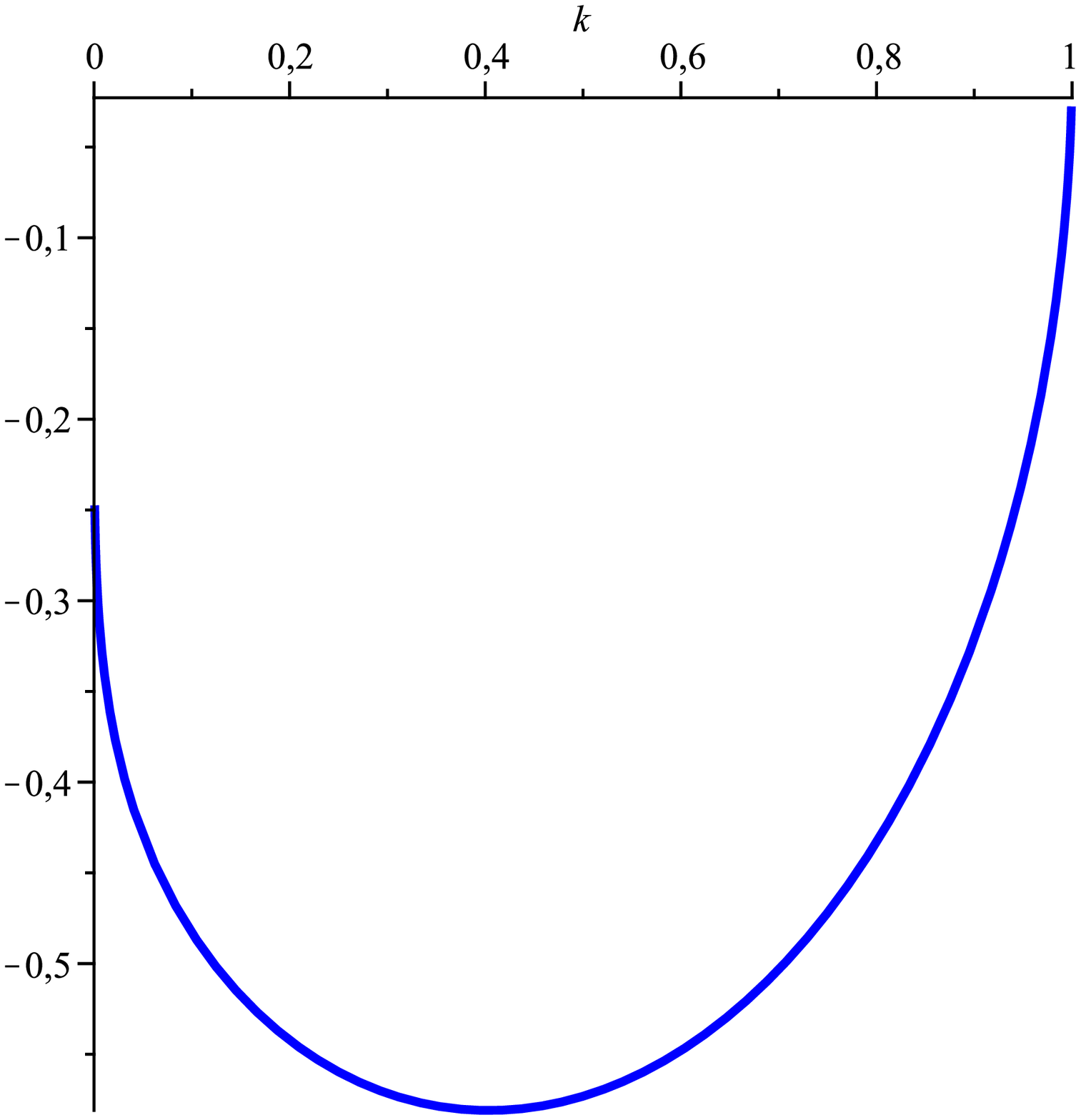}B.$\,\,\,\,\,\,\,\,\,\,\,\,$
\includegraphics[width=35mm]{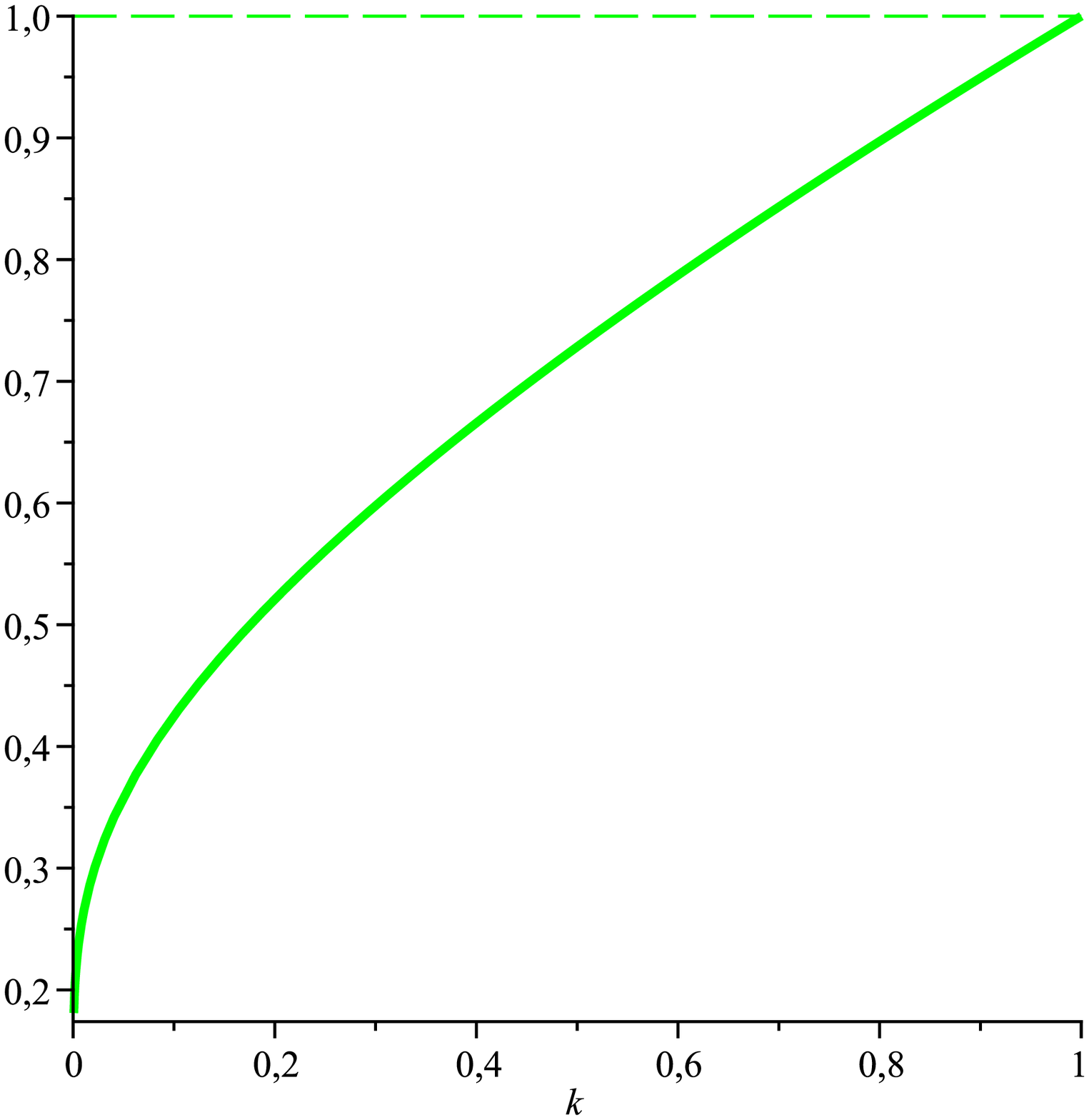}.C
\caption{
A. The estimate
(A.5) 
is shown. Along the horizontal axis time values $t$ are plotted (with $\Omega=1$),
and on vertical axis the relation $\frac{|r_1|}{d_1}$ is plotted.
B. The estimate (A.12) 
is shown. Along the horizontal axis time values $k$ are plotted,
and on vertical axis $L(k)$ given by the relation
(A.12) 
is plotted.
C.
The estimate (A.13) 
is shown. Along the horizontal axis time values $k$ are plotted,
and on vertical axis $\dfrac{\pi}{2{\rm {\bf{K'}}}}$ is plotted.}
\label{F:estimation-delta2-m}
\end{figure}

Using explicit estimates (see. Fig. \ref{F:estimation-delta2-m}.B and
Fig.\ref{F:estimation-delta2-m}.C, respectively)
\bea
L(k)\equiv\exp\left(-\dfrac{\pi {\rm {\bf {K}}}}{2{\rm {\bf
{K'}}}}\right)-\sqrt{1-k^2}&\leq & 0
\label{est-exp-f-m},\\
\dfrac{\pi}{2{\rm {\bf {K'}}}}&\leq &1 \label{est-exp-k-m}; \eea we
have \bea\exp\left(-\dfrac{\pi {\rm {\bf {K}}}}{2{\rm {\bf
{K'}}}}\right)\exp\left(\dfrac{\pi\Omega t}{2{\rm {\bf
{K'}}}}\right)<k'e^{\Omega t}. \eea

We require that
\be \label{cond-m}
     \biggl(k'\biggr)^2\exp\left(2\Omega t\right) < \frac12
\ee
to estimate (\ref{RHS-m}).

Then we have \bea d_1\leq 2  \dfrac{\pi}{k {\rm {\bf {K}}}'}
\biggl(k'\biggr)^3e^{3\Omega t}< \dfrac{4}{k }
\biggl(k'\biggr)^3e^{3\Omega t} \label{RHSm-m}.\eea

Notice that for $k'<1/2$ \be \label{Omega-k} \Omega =\mu
\frac{1}{\sqrt{1-2k^{\prime 2}}}\leq \mu \sqrt 2\ee
and (\ref{cond-m})
are true, if
 \bea\label{den}
\biggl(k'\biggr)^2\exp\left(2\Omega t\right)\leq
\biggl(k'\biggr)^2\exp\left(2\sqrt 2\mu t\right)\leq
\frac12,\,\,\,\,k'\exp\left(\sqrt 2\mu t\right)\leq \frac1{\sqrt2}.
\eea
Consequently
(\ref{den}) is true, for
\be \label{t-rest} t<\frac{c}{\mu\sqrt 2
}\ln 1/k', \,\,\,\,c<1/2.
\ee
Indeed, \be k'\exp\left(\sqrt 2\mu
t\right)<(k')^{1-c}<1/\sqrt{2}\,\,\,\, {\rm if }\,\,\,\,c<1/2,\,\,k'<1/2.\ee

With (\ref{t-rest}) the following is true \bea \label{d1} d_1\leq \dfrac{4}{k }
\biggl(k'\biggr)^3e^{3\Omega t}<\dfrac{4}{k }(k')^{3(1-c)}. \eea

In the same way for \bea
 r _n\equiv\dfrac{\pi}{k {\rm {\bf
{K}}}'} \left\{-\dfrac{\sinh ((2n+1)\rho' u'')}{\cosh ( (2n+1)\rho
')}+ \,\dfrac{\sinh ( (2n+3)\rho' u'')}{\cosh ( (2n+3)\rho
')}+...\right\} \eea we have \be
 |r _n|
\leq d_n,\label{main_estmn} \ee where \be d_n\equiv\dfrac{\pi}{k {\rm
{\bf {K}}}'} \dfrac{e^{-(2n+1)\rho'(1 - u'')}} {1 - e^{-2\rho'(1 -
u'')}} \label{main-d} \ee and \bea \label{est-dn} d_n\leq \dfrac{4}{k
} \biggl(k'\biggr)^{2n+1}e^{(2n+1)\Omega t}<\dfrac{4}{k
}(k')^{(2n+1)(1-c)}. \eea

For small $k'<1/2$ we get $k>1/2$ and

\bea d_1\leq
\dfrac{4}{k } \biggl(k'\biggr)^3e^{3\Omega t}\leq
\dfrac{8}{k } \biggl(k'\biggr)^3e^{3\Omega t}. \eea

Finally for  $k'<1/2$  $k>1/2$ we get
\bea d_1\leq
\dfrac{4}{k } \biggl(k'\biggr)^3e^{3\Omega t}\leq 8
\biggl(k'\biggr)^3e^{3\Omega t}.\eea

\section{Bounds of $\Delta$}
Having introduced notations
\be
\Delta_{n}=\mu t\delta_n,
\ee
we obtain:
\bea
\label{delta-2} \delta_0&=&1-
 \dfrac{\pi }
         {2{\rm {\bf K'}}} \frac{1 }{\sqrt{1-2k^{\prime 2}}} \\\label{delta-4}
\delta_1&=&\biggl(1+\frac{3A^2}{8\mu^2}\epsilon \biggr)-
 \dfrac{\pi }
         {2{\rm {\bf K'}}} \frac{1 }{\sqrt{1-2k^{\prime 2}}} \\\nonumber&=&1+
 \frac38 \left(1+\frac{1}{1-2k^{\prime 2}}\right)
\left(\dfrac{\frac{\pi}{k {\rm {\bf {K}}}'}}{\cosh \biggl( \dfrac{\pi {\rm
{\bf K}}}{2{\rm {\bf K'}}}\biggr)}\right)^2 -
 \dfrac{\pi }
         {2{\rm {\bf K'}}} \frac{1 }{\sqrt{1-2k^{\prime 2}}}      \\
\label{delta-6}
 \delta_2&=&1+\frac{3A^2}{8\mu^2}\epsilon- \dfrac{15 A^4}{2^8 \mu^4}\epsilon^2 -
 \dfrac{\pi }
         {2{\rm {\bf K'}}} \frac{1 }{\sqrt{1-2k^{\prime 2}}}  \\
        &=& 1+\frac{3}{8}\left(1+\frac{1}{1-2k^{\prime 2}}\right)
\left(\dfrac{\frac{\pi}{k {\rm {\bf {K}}}'}}{\cosh \biggl( \dfrac{\pi {\rm
{\bf K}}}{2{\rm {\bf K'}}}\biggr)}\right)^2 - \frac{15 }{2^8
}\left(1+\frac{1}{1-2k^{\prime 2}}\right) \left(\dfrac{\frac{\pi}{k
{\rm {\bf {K}}}'}}{\cosh \biggl( \dfrac{\pi {\rm {\bf K}}}{2{\rm {\bf
K'}}}\biggr)}\right)^4 \nonumber \\ &-& \dfrac{\pi } {2{\rm {\bf K'}}} \frac{1 }{\sqrt{1-2k^{\prime 2}}}
         \nonumber
\eea
     These formulae are based on the relations:
     \bea
     \label{Omega-k}
    \dfrac{\pi \Omega }
         {2{\rm {\bf K'}}}&=&\dfrac{\pi }
         {2{\rm {\bf K'}}}
           \frac{\mu }{\sqrt{1-2k^{\prime 2}}}\\
\label{A-a} \frac{A^2}{\mu^2}\epsilon
&=&\left(1+\frac{1}{1-2k^{\prime 2}}\right)
\left(\dfrac{\frac{\pi}{k {\rm {\bf {K}}}'}}{\cosh \biggl( \dfrac{\pi {\rm
{\bf K}}}{2{\rm {\bf K'}}}\biggr)}\right)^2 \eea

We have (see Fig. \ref{mu-omega-m})
       \be
       \label{mu-Omega}
      \delta _0<\sqrt{2}k^{\prime 2} \,\,\,\,\,\,\,,0\leq k^{\prime }\leq
      \frac12\ee
         and, consequently,
            \be
       \label{mu-Omega-m}
    \Delta_0<\sqrt{2}\mu k^{\prime 2}t \,\,\,\,\,\,\,,0
    \leq k^{\prime }\leq \frac12.\ee
We also have
\be \delta_1< k^{\prime 4}.
\label{delta-4m}
\ee
This estimation is illustrated in  Fig.\ref{delta-4-6}.A.
\begin{figure}[h!]
\centering
\includegraphics[width=45mm]{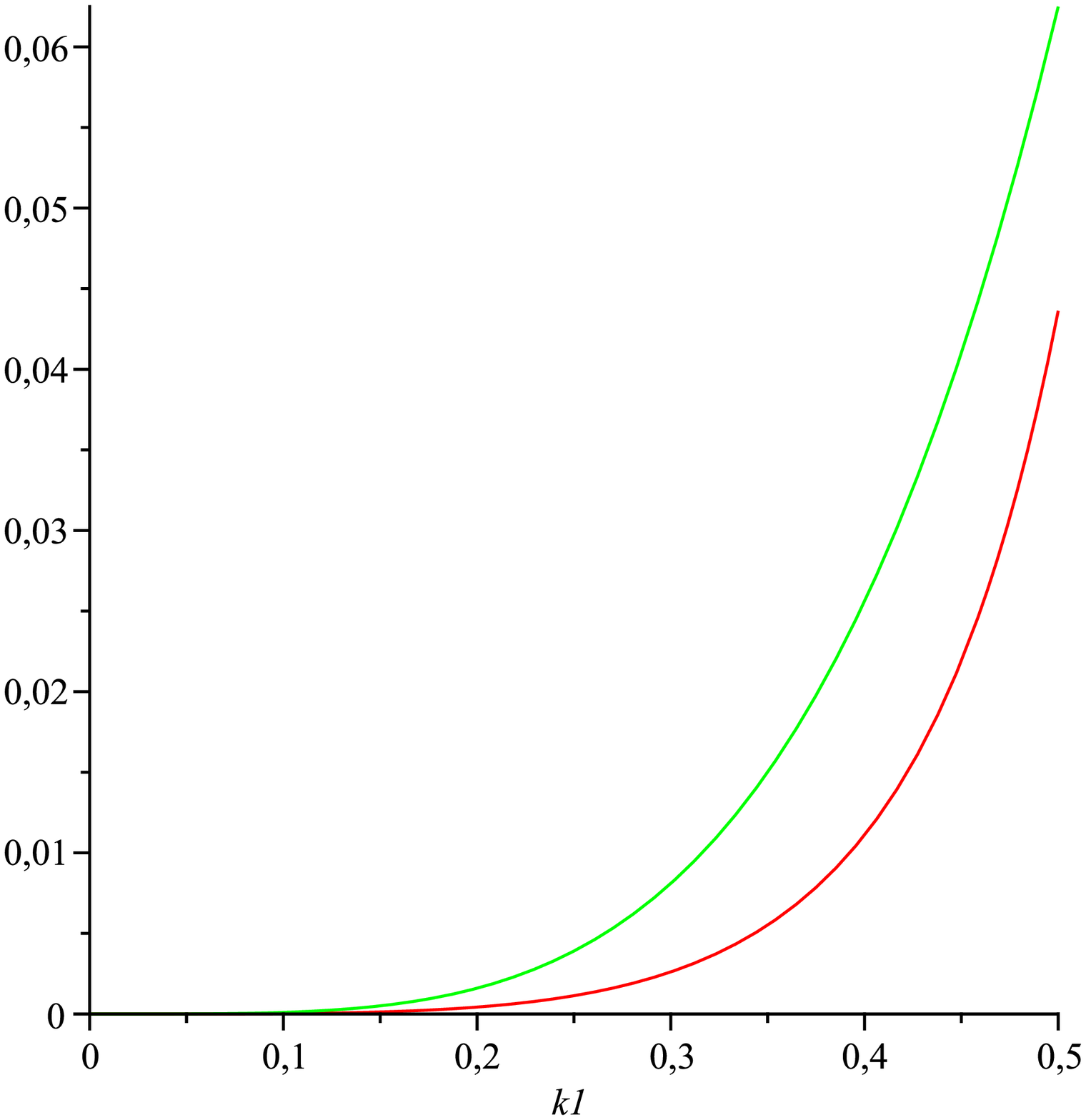}$A\,\,\,\,\,\,\,\,\,\,\,\,\,\,\,\,\,\,\,\,$
\includegraphics[width=45mm]{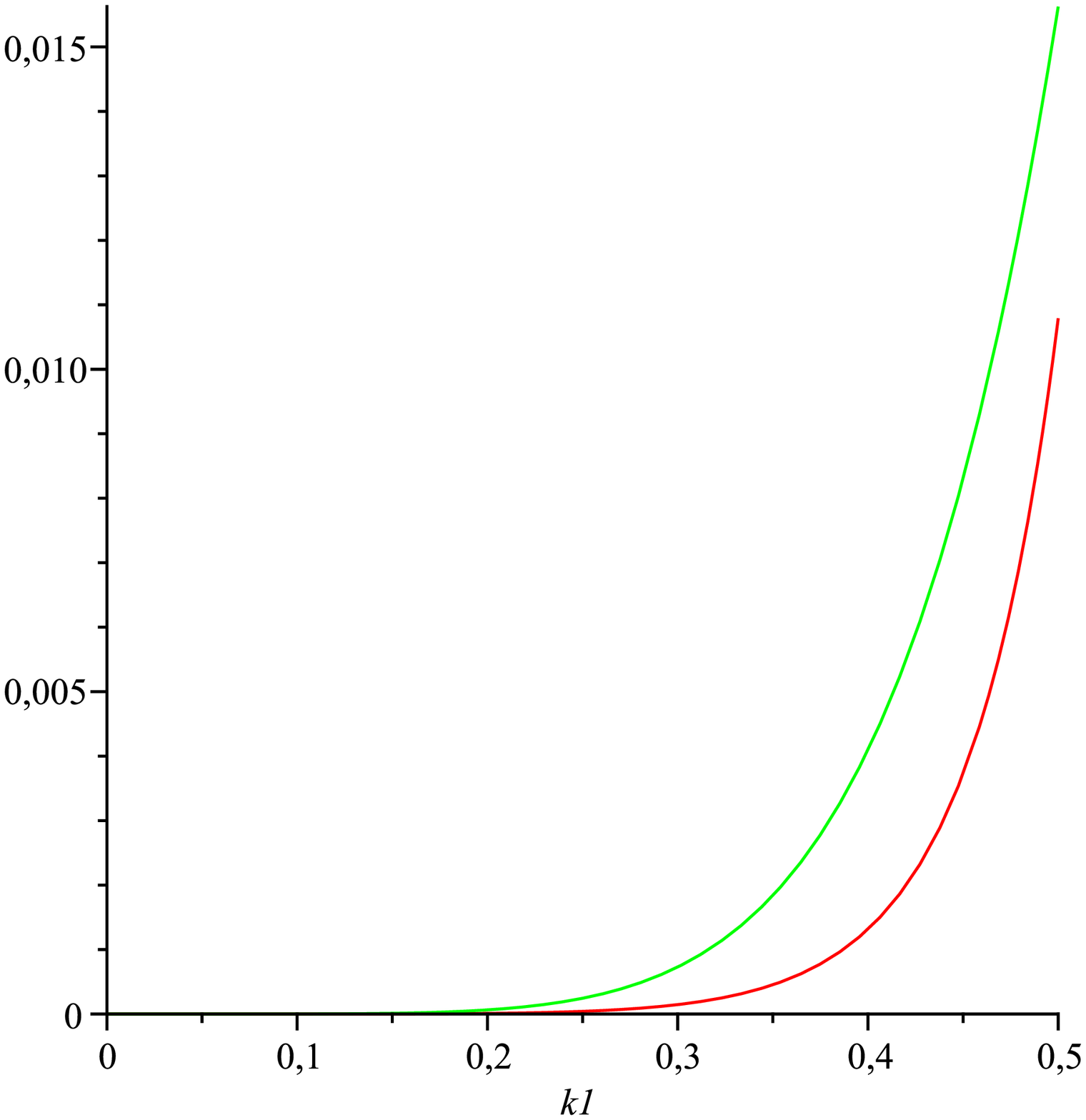}$B$
\caption{ (color online)
A. The plot illustrates the estimation
(B.9). 
On the horizontal axis $k^{\prime 2}$
is plotted, on the vertical axis the values of  the left-hand side of
(B.3) 
(red color) and  the function $k^{\prime 4}$ (green color) are presented.
B. The plot illustrates the estimation
(B.4). 
 On the horizontal axis $k^{\prime 2}$
is plotted, on the vertical axis the values of  the left-hand side of
(B.4) 
(red color) and  the function $k^{\prime 6}$ (green color) are presented.}
\label{delta-4-6}
\end{figure}

From (\ref{delta-4m}) we get that
 \be \Delta_1\leq \mu
k^{\prime 4}t. \ee

In Fig.\ref{delta-4-6}.B. we show functions  $\delta _2=\delta _2(k^{\prime })$ and
$k^{\prime 6}$. From from this plot one can see that
\be
\delta_2<k^{\prime 6}\ee and \bea \Delta_2&\leq&\mu
k^{\prime 6}t.
         \eea

\end{document}